\pdfoutput=1
\documentclass[12pt]{article}
\def\baselinestretch{1.175} 

\usepackage{lipsum}

\usepackage{setspace}
\usepackage{latexsym} 
\usepackage{verbatim}
\usepackage{tikz}
\usetikzlibrary{matrix}
\usepackage{color}
\usepackage{graphicx,amssymb,amsfonts,amsmath,amssymb,amscd,amstext, mathrsfs}
\usepackage{graphicx} 
\usepackage{bbm}
\usepackage{dsfont}
\usepackage{xcolor}
\definecolor{darkblue}{RGB}{0,0,160}

\usepackage[
    colorlinks=true,
    citecolor=darkblue,
    linkcolor=black,
    urlcolor=darkblue
]{hyperref}

\usepackage{amsmath}
\allowdisplaybreaks

\usepackage{fancyhdr}
\usepackage{empheq}
\usepackage{hyperref}

\newlength\dlf

\def\be{\begin{eqnarray}}
\def\ee{\end{eqnarray}}
\def \bea {\begin{equation}}
\def \eea {\end{equation}}

\def \nn {\nonumber}

\def \d {\partial}

\def \rr {\raise.35ex\hbox{\small $\prime$}\kern-.17em{\mbox{\large $\imath$}}}

\def \dels {\partial\kern-.5em / \kern.5em}
\def \As {{A\kern-.5em / \kern.5em}}
\def \Ds {D\kern-.7em / \kern.5em}

\def \z {(0)}

\def\frac#1#2{{#1\over #2}}

\def \zb {\bar z}
\def\dtb {\partial_{\bar t}}

\def \iffa {\iffalse} 

\def \ed  {\end{document}}

\def\nn{\nonumber}

\def \ed  {\end{document}}

\DeclareFontShape{OT1}{cmr}{mx}{n}{<->cmr10}{}
\DeclareFontShape{OT1}{cmr}{mx}{b}{<->cmbx10}{}

\usepackage{pdfrender}
\newcommand{\titlefontmedium}{%
  \fontfamily{cmr}\fontseries{mx}\selectfont
  \pdfrender{TextRenderingMode=FillStroke,LineWidth=0.375pt}
}

\def \d_t{\partial_t}

\def \A{{\cal A}}

\def \z{\bar z}

\usepackage{cite}

\begin{document}

\begin{titlepage}
\begin{flushright} 
\end{flushright}
\begin{center} 
\vspace{2cm}  

{\fontsize{18pt}{0pt}{\selectfont\titlefontmedium
A Matrix View  of  Holographic Null-State Equations in $d\geq 2$ 
}} 
\vspace{2.75cm}  
\\
 {\fontsize{11pt}{0pt}{Kuo-Wei Huang
}}   
\\ 
\vspace{1cm} 
{\fontsize{11pt}{0pt}{$^1$\it Department of Physics, National Cheng Kung University (NCKU), 
Tainan, Taiwan}}\\
\vspace{0.1cm} 
{\fontsize{11pt}{0pt}{$^2$\it  Physics Division, National Center for Theoretical Sciences, 
Taipei, Taiwan}}
\end{center}
\vspace{1.5cm} 

\begin{center} 
{\selectfont\titlefontmedium  Abstract}
\end{center} 
{\noindent 
We show that the null-state-type partial differential equations recently derived from AdS gravity with a black hole can be efficiently organized using finite-dimensional matrices with systematic off-diagonal structures. This matrix formulation generates an infinite tower of CFT differential equations at arbitrary levels. The $d=2$ matrix reproduces a known result based on Virasoro null-state vectors at large central charge and admits an algebraic representation in terms of a linear combination of the identity matrix and $\mathfrak{sl}_2$ generators, with a heavy-state deformation encoding the central charge contribution. The $d=4$ governing matrix exhibits a similar yet richer structure involving products of $\mathfrak{sl}_2$ generators. We also discuss recursion relations and factorized product structures associated with these holographic differential equations.
}

\end{titlepage}

\addtolength{\parskip}{1 ex}
\jot=1 ex

\setcounter{page}{2}
\tableofcontents

\section{{\selectfont\titlefontmedium Introduction}}

A main motivation for this work comes from the enduring hope of extending the remarkable successes of 
conformal field theory (CFT) in two spacetime dimensions to higher dimensions -- a possibility already 
mentioned by A. Polyakov \cite{Polyakov:2003interview}:
\vspace{0.1cm}
{\begin{spacing}{1.05}
\textit{``We can classify possible fixed points in two dimensions using operator product expansions. 
That’s what conformal field theories are about. And I think it’s not excluded, that in three dimensions 
something like that is still possible." 
}
\end{spacing}}
\vspace{0.1cm} 

Inspired by this challenge, and by recent fruitful progress in computing conformal observables 
using techniques from holographic duality \cite{Maldacena:1997re, Gubser:1998bc, Witten:1998qj} 
and the conformal bootstrap (see, $e.g.$, \cite{Poland:2018epd} for a review), In the present work, we aim to make a modest contribution toward the broader goal of advancing the analytic study of higher-dimensional quantum field theories at strong coupling. More specifically, we seek to identify compact organizing structures 
in higher-dimensional CFTs that echo those found in two-dimensional CFTs.   
As a concrete and tractable setting, we focus on a class of four-dimensional holographic CFTs.  
Our analysis is confined to the holographic setting and does not address the classification of fixed points itself.  
We hope, however, that the new results will help organize the analytic structures 
of higher-dimensional conformal observables in a compact and systematic manner.

Since the seminal work \cite{BELAVIN1984333}, it has been well understood that the stress-tensor 
operator product expansion (OPE) in two-dimensional CFTs gives rise to the Virasoro algebra and its representation theory.  
In particular, degenerate primary operators have null descendants, whose decoupling leads to linear partial differential equations for correlation functions involving these operators. 
These differential equations provide powerful analytic tools in two-dimensional CFTs. In higher dimensions, the stress-tensor OPE has a considerably richer structure, as reflected, for example, in the stress-tensor three-point function \cite{Osborn:1993cr}. Rather than working directly with the higher-dimensional OPE, we adopt a holographic approach to identify null-state-type differential equations for the dual CFT correlators. We then view these equations as a potential guide to uncovering the organizing principles of higher-dimensional CFTs.

The conformal observable we study is a scalar four-point function. By conformal transformations, its four operator insertions can be placed on a two-dimensional plane. This kinematic simplification motivates the question of whether certain organizing structures familiar from two-dimensional CFTs have analogues in higher dimensions, even when the underlying theory remains genuinely higher-dimensional. In holographic CFTs, this question can be naturally connected to a thermal scalar two-point correlator, computed by solving a bulk scalar field equation in a black hole background. 
In the dual CFTs, the thermal two-point correlator can be interpreted as a heavy-light four-point function at large central charge, in which two heavy scalars of parametrically large conformal dimension create the thermal background.  

We should clarify which operator exchanges contribute to the correlator and which are captured by the differential equations studied here.
In CFTs with large central charge (large N),  the correlator receives contributions from double-trace operators constructed from pairs of scalar operators, as well as from composite multi-stress tensor primaries built from the stress tensor and its derivatives. Generalizing the two-dimensional Virasoro null-state equations, we seek analogous differential equations in higher dimensions whose solutions capture and isolate the multi-stress tensor contributions. Including the identity contribution, the resummed stress-tensor sector provides a higher-dimensional analogue of the two-dimensional Virasoro vacuum block.  
For related discussions of Virasoro blocks in two-dimensional CFTs, see, $e.g.$, \cite{zamolodchikov1984conformal, zamolodchikov2, Harlow:2011ny, Hartman:2013mia, Fitzpatrick:2014vua, Perlmutter:2015iya, Hijano:2015qja, Fitzpatrick:2015foa, Beccaria:2015shq, Fitzpatrick:2015dlt, Fitzpatrick:2016ive, Chen:2016cms, Fitzpatrick:2016mtp, Hikida:2018dxe, Bombini:2018jrg, Cotler:2019nbi, Besken:2019jyw, Anous:2020vtw, Nguyen:2021jja, Pavlov:2024zqj, Bissi:2024wur, Fortin:2025asz}.

A further motivation for this work is to develop analytic tools for probing black hole interiors through boundary dual CFTs, particularly in higher spacetime dimensions.   Based on a holographic computation, a recent work \cite{Ceplak:2024bja} showed that a complexified trajectory reaching the central singularity of a classical AdS black hole (the so-called bouncing singularity) is encoded in the stress-tensor sector of the thermal scalar two-point function in the boundary dual CFTs. Earlier studies include, $e.g$, \cite{Kraus:2002iv, Fidkowski:2003nf, Festuccia:2005pi}.  
The structure of finite-temperature conformal correlators and their connection to black hole interiors have attracted growing interest and have been investigated using complementary approaches based on dual gravity and field-theoretic methods; see, $e.g.$, \cite{Iliesiu:2018fao, Gobeil:2018fzy, Kulaxizi:2018dxo, Fitzpatrick:2019zqz, Li:2019tpf, Kulaxizi:2019tkd, Fitzpatrick:2019efk, 
Karlsson:2019dbd, Li:2019zba, Karlsson:2019txu, Karlsson:2020ghx, Fitzpatrick:2020yjb, Alday:2020eua, Grinberg:2020fdj, Karlsson:2021duj, Rodriguez-Gomez:2021mkk, Krishna:2021fus, Karlsson:2022osn, Dodelson:2022yvn,  Parisini:2022wkb, Dodelson:2023vrw, Esper:2023jeq, Parisini:2023nbd,  Buric:2025anb, Haehl:2025ehf,  Buric:2025uqt, Ammon:2025cdz, Buric:2025fye, Niarchos:2025cdg, Bajc:2025jjv,  Barrat:2025twb, Afkhami-Jeddi:2025wra, Dodelson:2025jff, Ceplak:2025dds,  Jia:2026pmv,   Araya:2026shz, Giombi:2026kdz, Jia:2026ryl,   Arnaudo:2026der, Grozdanov:2026cut,  Arnaudo:2026tcy,Grozdanov:2026ktq,  Niarchos:2026wsw, Arnaudo:2026axe,  Hartnoll:2026vhu, Ceplak:2026ard,  Guo:2026xyl, Komargodski:2026oob, Batra:2026gbq, Correia:2026fnr,  Buric:2026sit, Bajc:2026wzi,   Chakravarty:2026fdm,  Chen:2026efa,    Almheiri:2026cpv, Oh:2026dky, Aniceto:2026vtv, Grassi:2026ekz}.  The higher-dimensional CFT differential equations studied here are closely related to these developments: these equations govern the resummation of multi-stress tensor exchanges in the thermal scalar two-point function. The goal of the present work, however, is more specific: we would like to find a compact way to organize these null-state-type equations to all levels, in both two- and four-dimensional holographic CFTs, leaving their analytic solutions and potential implications for black hole interiors to future work. 

\subsection*{\selectfont\titlefontmedium Summary of the Main Results}

The present work builds on the recent holographic derivation of four-dimensional null-state-type differential equations \cite{Huang:2026jbv}, obtained by studying a light probe massive scalar minimally coupled to gravity in an AdS black hole background.  At certain special values of the scalar's conformal dimension, the near-boundary expansion of the bulk field equation exhibits a truncation that gives rise to partial differential equations for the conformal correlator. This holographic analysis confirms previously conjectured three equations governing the resummed contributions of minimal-twist multi-stress tensor operators \cite{Huang:2023ikg}, and also establishes additional differential equations, including one that extends beyond the near-lightcone regime. See also \cite{Huang:2024wbq} for a discussion of solutions to this type of partial differential equations. 

Here we develop a matrix formulation that organizes the null-state-type equations at arbitrary levels, in both two- and four-dimensional holographic CFTs. Our approach provides a systematic construction of an infinite tower of partial differential equations for the degenerate-type scalars, without requiring a lightcone limit. We now highlight and summarize some of the main results. 

\subsubsection*{$\bullet$ \selectfont\titlefontmedium $d=2$ Holographic CFT}

In two dimensions, we focus on the holomorphic sector and express the thermal scalar two-point function at large central charge as a heavy-light four-point correlator, which we denote by 
\begin{equation}
\Phi_0(z)=\langle {\cal O}_H(\infty){\cal O}_L(z){\cal O}_L(0){\cal O}_H(1)\rangle \ . 
\end{equation}  
In the holographic description, we also use $\Phi_0$ to denote the leading term in the near-boundary expansion of the bulk-to-boundary propagator, which satisfies the bulk scalar field equation in a black hole background.
Applying the Belavin-Polyakov-Zamolodchikov (BPZ) level-2 null-state differential equation \cite{BELAVIN1984333} to this four-point function gives
\begin{align}
\Big(\partial_z^2+\frac{6\eta}{(1-z)^2}\Big) \Phi_0(z)\Big|_{\Delta= -1 }=0 
\end{align}
where $\eta= {h_H\over c}$, held fixed at large central charge, encodes the contribution of the heavy operators. 
This null-state differential equation requires the light scalars to have holomorphic weight $h_L= - {1\over 2}+{\cal O}({1\over c})$. We write $\Delta= 2h_L$, and identify $n$ as the level number, so the level-$2$ equation requires $\Delta= -1$ at large $c$.

It is convenient to introduce a new coordinate $t$ defined by
\begin{align}
&z = 1- e^{-t}  \ .
\end{align} 
We then write the correlator $\Phi_0(z)$ as $V(t)$. From the holographic computation, detailed in the main text, we find that the level-$n$ null-state-type equation can be written as follows: 
\begin{align}
\mathscr{D}_n V(t)\Big|_{\Delta= 1-n} =0 \ , ~~~~~~~~ \mathscr{D}_n = \det { M}_{n} 
\end{align}  
where $M_n$ is an $n\times n$ matrix that involves the differential operator $\d_t$:
\begin{align}
\label{2dMintro}
({M}_n)_{ij} =
\begin{cases}
\d_t+i-1 ~~~~~~ & i=j \\
1 ~~~~~~ & i=j-1 \\
-6 \eta (n-j) j   ~~~~~~ & i=j+1 \\
0 ~~~~~~ &\text{\it otherwise}   
\end{cases} 
\end{align}  
The matrix construction efficiently generates differential equations at arbitrary integer levels $n\geq2$, with corresponding degenerate light scalar dimensions given by $\Delta=1-n$.

The $d=2$ organizing matrix \eqref{2dMintro} admits a decomposition in terms of generators of $\mathfrak{sl}_2$:
\begin{align}
{M}=\Big(\d_t+\frac{n-1}{2}\Big) { I} +  { J}_0 + {J}_{-} - 6 \eta {J_+} 
\end{align}
where $I$ denotes the $n\times n$ identity matrix, and we choose the generators
\begin{align}
(J^{(n)}_0)_{ij} =  \Big(i-{n+1\over 2}\Big)\delta_{ij} \ , ~~~~~~~  
(J^{(n)}_+)_{ij} = j\big(n-j\big)\delta_{i-1, j} \ ,   ~~~~~~~ 
(J^{(n)}_-)_{ij} =\delta_{i+1, j} 
\end{align} with $i,j=1,\ldots,n$. 
They satisfy the commutation relations $[J_0,J_\pm]=\pm J_\pm$, $[J_+,J_-]= 2J_0$.  
From the dual gravity perspective, $\eta=0$ corresponds to the absence of a black hole, and in this case, the matrix is undeformed by the heavy-state background.

Diagonalizing $M_n$, we obtain a product formula for its determinant:
\begin{align}
\label{2dPD}
 { \mathscr{ D}}_n = \det M_n= 
\prod_{k=-(n-1)}^{n-1}
\left(
\d_t+\frac{n-1}{2}
+{k\over 2}\sqrt{1-24 \eta}
\right)   \ .
\end{align}  Derived here from AdS$_3$ gravity, this expression for the null-state differential operator agrees with the known result obtained from the CFT formalism based on Virasoro representation theory \cite{Bauer:1991ai}, after taking the large central charge limit The corresponding $d=2$ CFT result appears in Eq.~(3.9) of \cite{Chen:2016cms}; see also Appendix A of \cite{Fitzpatrick:2016ive} and Exercise 8.8 of \cite{DiFrancesco:1997nk} for related discussions.

\subsubsection*{$\bullet$ \selectfont\titlefontmedium $d=4$ Holographic CFT}

In the AdS$_5$ gravity computation, we consider a minimally-coupled bulk scalar in a Schwarzschild-AdS black hole background with a spherical horizon. 
We expand the corresponding bulk-to-boundary propagator $\Phi$ near the AdS boundary and denote its leading coefficient by $\Phi_0$.
To simplify the near-boundary expressions, in this case we find it useful to let
 \begin{align}
\Phi_{0}(z,\zb) = \frac{1-\bar z}{z-\bar z}\, Q  (z,\zb)    \ . 
\end{align}   
The first $d=4$ differential equation requires $\Delta= -1$. More generally, we write the degenerate light scalar dimension as $\Delta=2-n$, with $n\geq 3$, in the $d=4$ analysis.

Three null-state-type differential equations in $d=4$ were first written down by guesswork in   \cite{Huang:2023ikg}, focusing on the simplifying lightcone limit, and were supported by several nontrivial checks using minimal-twist multi-stress tensor OPE data.   
Subsequently, these CFT equations were derived from holography in \cite{Huang:2026jbv}. In particular, this latter work also obtains the full lowest-level ($n=3$) equation without taking the lightcone limit. The corresponding equation reads 
 \begin{align}
\Big[ \, {\cal D}^{(3)}  {\cal \overline D}^{(3)}  - 3 {\cal D}^{(3)}  {\cal \overline D}^{(2)}  
+ \mu \big({\cal D}^{(2)}   +  {\cal \overline D}^{(2)} + {\cal D}  {\cal \overline D} + {\cal D} \big)\, \Big] \, Q(z,\zb)  \Big|_{\Delta = -1} = 0  
\end{align} 
where we have introduced the shorthand notation
 \begin{align}
{\cal D}^{(n)}\equiv (1-z)^n \partial^n  \ , ~~~~~ {\cal \overline D}^{(n)}\equiv (1-\bar z)^n \bar\partial^n 
\end{align}  
for the weighted differential operators. The parameter $\mu$ is the ratio
of the conformal dimension of the heavy scalar to the $d=4$ CFT central charge $C_T$.

We introduce a change of coordinates involving both $z$ and $\bar z$:
\begin{align}
\label{ztnotation}
&z = 1- e^{-t} \ , ~~~ \bar z= 1- e^{-\bar t} \ .
\end{align}  
In the $(t, \bar t)$ coordinates, we write the correlator $Q(z,\bar z)$ by $V(t,\bar t)$.  
The holographic computation yields the following matrix form of the level-$n$ null-state-type differential equation in $d=4$: 
\begin{align}
\mathscr{D}_n V(t,\bar t)\big|_{\Delta=2-n}=0\ , ~~~~~ \mathscr{D}_n= \det {M_n} \ , ~~~~~ (M_n)_{ij}
=\begin{cases}
(\d_t+i-1)(\partial_{\bar t}+i-2) ~~~~~& i=j \\[3pt]
i \, \big(\d_t+\partial_{\bar t}+2(i-1)\big) ~~~~~& i=j-1\\[3pt]
i \, (i+1) & i=j-2\\[3pt]
\dfrac{\mu}{4}(n-i+1) (n-i+2)  ~~~~~& i=j+2\\[3pt]
0 ~~~~~ & \it{otherwise} 
\end{cases}\nn\\ 
\end{align} 
where $M_{(n)}$ is an $n\times n$ matrix whose entries involve $\d_t$ and $\partial_{\bar t}$. This $d=4$ matrix generates the differential
equations at arbitrary levels, without taking a lightcone limit. 

We find that the $d=4$ governing matrix can be expressed in terms
of products of $\mathfrak{sl}_2$ generators:
\begin{align}
M=
\left[\left(\d_t+\frac{1}{2}\right)I+\A \, \right]&
\left[\left(\partial_{\bar t}-\frac{1}{2}\right)I+\A \, \right]
+\frac{\mu}{4} \, J_+^2  \ , ~~~~~
\A=\frac{n-2}{2}\, I+J_0-J_- \ . 
\end{align}  
Here $I$ denotes the $n\times n$ identity matrix and the generators are chosen to be 
\begin{align}
(J_0)_{ij}=\left(i-\frac{n+1}{2}\right)\delta_{ij} \ , ~~~~~
(J_+)_{ij}=(j-n)\delta_{i,j+1} \ , ~~~~~
 (J_-)_{ij}= -i\,\delta_{i,j-1} 
\end{align}  
with $i,j=1,\ldots,n$.  
They satisfy the commutation relations $[J_0,J_\pm]=\pm J_\pm$, $[J_+,J_-]= 2J_0$.

In addition, we introduce the following commuting differential operators:
\begin{align}
\widehat{\alpha}=\frac{\d_t-\partial_{\bar t}+1}{2} \ ,
\qquad
\widehat{s}=\frac{\d_t+\partial_{\bar t}-1}{2} \ . 
\end{align}
The $d=4$ matrix can be written as $M_n=B_n(\widehat{s})-\widehat{\alpha}^{\,2}I_n$.
Let $\lambda_j(\widehat{s})$ denote the eigenvalues of the
$n\times n$ matrix $B_n$. The determinant consequently
takes a product form
\begin{align}
\mathscr{D}_n
=\det M_n
=\prod_{j=1}^{n}
\left[-\Big(\frac{\d_t-\dtb+1}{2}\Big)^{2}+\lambda_j(\widehat{s})\right] \ .
\end{align}  
Unlike its $d=2$ counterpart, the $d=4$ matrix involves quadratic
combinations of the $\mathfrak{sl}_2$ generators, rather than a linear
combination of them. Although the $d=4$ matrix can still be  diagonalized and  
its eigenvalues can be expressed
in terms of algebraic roots, this does not lead to a simple 
product form analogous to \eqref{2dPD}. Nevertheless, the matrix
provides a systematic algorithm for constructing $d=4$
null-state-type differential equations at arbitrary levels in the holographic CFTs.

\subsection*{\selectfont\titlefontmedium Outline}

In section \ref{sec:2d}, we first review the holographic derivation of the
level-2 null-state equation in $d=2$ holographic CFTs. We then develop a
matrix formulation that organizes the differential equations at arbitrary
levels. We derive a recursion relation between matrices at different
levels, obtain a related combinatorial expression, and
establish the factorized product formula for the determinant. We also discuss
the algebraic structure underlying the $d=2$ matrix.

 We turn to $d=4$ holographic CFTs in section \ref{sec:4d}. After reviewing
the recent holographic derivation of a null-state-type equation from
AdS$_5$ gravity, we find a compact matrix organization of the
differential equations at arbitrary levels and derive the corresponding
recursion relation. We obtain an algebraic representation of the
$d=4$ matrix and compare its structure with that of the $d=2$ matrix.
We also discuss the corresponding matrix relevant to the $d=4$ scalar correlator in the lightcone limit,  
where minimal-twist operators dominate. We conclude with some future questions in section~\ref{sec:dis}.

\section{\selectfont\titlefontmedium Two-Dimensional Holographic CFT}
\label{sec:2d}

\subsection{\selectfont\titlefontmedium Review: Level-2 BPZ Differential Equation from AdS$_3$ Black Hole}

We begin with the simplest example, following \cite{Huang:2026jbv}, in
which the level-2 BPZ null-state differential equation at large central charge is
derived by studying a massive light probe scalar field in an AdS$_3$ spherical black hole
background.

Setting the AdS radius to unity, we write the black hole metric as
\begin{align}
\label{AdS3matric}
ds^2 = \left(1+r^2f(r)\right)dt^2+\frac{dr^2}{1+r^2f(r)}+r^2d\theta^2 \ , ~~~~~ f(r)=1-\frac{\mu}{r^2} \ . 
\end{align}
The parameter $\mu$ is proportional to the black hole mass and, through holography, is related to the ratio of the heavy scalar's conformal weight to the central charge $c$.   
We adopt the convention
\begin{align}
\mu=24\, \eta \ , \qquad \eta \equiv \frac{h_H}{c}  \ .
\end{align} 
In the heavy-light limit, both $h_H$ and $c$ are taken to infinity with the ratio $\eta$ held fixed.
Considering a minimally-coupled bulk scalar, the bulk-to-boundary propagator obeys the bulk field equation
\begin{align}
\left(-\nabla^2+m^2\right)\Phi(t,r,\theta)=0 \ . 
\end{align}
In the black hole background \eqref{AdS3matric}, this equation can be written as 
\begin{align}
\label{AdS3}
\Bigg[{1\over1+r^2-\mu}  \partial^2_t 
+{1\over r^2}\Big(\partial^2_\theta
+r(1+3r^2-\mu)\partial_r 
+r^2(1+r^2-\mu)\partial^2_r \Big)
\Bigg] \Phi=m^2\Phi  
\end{align} 
where the bulk scalar mass is related to the conformal dimension of its boundary dual via
\begin{align}
m^2=\Delta(\Delta-2) \ ,\qquad \Delta \equiv 2h \ .
\end{align}
The conformal correlator is given by the boundary limit\cite{Gubser:1998bc, Witten:1998qj}
\begin{align}
\lim_{r\to\infty}r^\Delta\Phi(t,r,\theta) \ .
\end{align}

Without a black hole, the pure AdS solution to the field
equation \eqref{AdS3} is given by \cite{Witten:1998qj}
\begin{align}
\label{pureAdS}
\Phi(t,r,\theta)\Big|_{\mu=0}=
\left(
{1\over2\left(\sqrt{1+r^2}\cosh t-r\cos\theta\right)}
\right)^\Delta  \ .
\end{align}
In the presence of a black hole, rather than using the exact AdS$_3$ 
bulk-to-boundary propagator obtained in \cite{Keski-Vakkuri:1998gmz},
we employ a near-boundary expansion since this approach can be straightforwardly
extended to higher dimensions. We write
\begin{align}
\Phi=r^{-\Delta} \left(
\Phi_0(t,\theta)
+\frac{\Phi_2(t,\theta)}{r^2}
+\frac{\Phi_4(t,\theta)}{r^4}
+\mathcal O(r^{-6}) \right) \ .
\end{align} 
In particular, as first observed in \cite{Fitzpatrick:2019zqz},
the near-boundary expansion of the bulk equation suffices to
capture the contributions from multi-stress tensors, without
specifying the bulk ($i.e.$, horizon) boundary condition. 

At leading order, the near-boundary expansion of the bulk equation
is automatically satisfied upon using the mass-dimension relation, $m^2 = (\Delta - 2)\Delta$. 
At subleading order, the equation is
 \begin{align}
\Big(\Delta^2 (\mu - 1) - \partial^2_t - \partial^2_\theta \Big) \Phi_{0} =  4 \Delta\,\Phi_{2}  \ .
\end{align}  
Decoupling $\Phi_2$ from this equation requires setting $\Delta=0$,
which, however, trivializes the analysis.
Instead, let us solve for $\Phi_2$ in terms of $\Phi_0$ and substitute
the result into the equation at the next order. This gives
 \begin{align}
\label{3Ddecouple}
\Big[(\partial_{t}^{2}+\partial_{\theta}^{2})^{2} + \alpha \partial_{t}^{2}+\beta \partial_{\theta}^{2}+\gamma \Big]\Phi_{0} = 32\,\Delta(\Delta+1)\,\Phi_{4} 
\end{align} 
where 
 \begin{align}
\begin{array}{cc}
\alpha =  2 \Delta(\Delta+4)(1-\mu) + 4(1-\mu) \ ,  & \\[-1mm]
 \beta =  2 \Delta(\Delta+2)(1-\mu) + 4 (1-\mu)  \ , &
\raisebox{0.5cm}{~~ $\gamma = \Delta^{2}(\Delta+2)^{2}(1-\mu)^{2} $ \ .}
\end{array}
\end{align} 
The coefficient of $\Phi_4$ in \eqref{3Ddecouple} vanishes at
\begin{align}
\Delta= -1 \ .
\end{align}  
This value, corresponding to $h= -\frac12$, is the large-$c$ limit of the weight of the level-2 BPZ degenerate operator.   
At this value, equation \eqref{3Ddecouple} becomes a differential equation for the CFT correlator. 

To express the equation in the standard $(z,\bar z)$ coordinates, consider the map
\begin{align}
\label{map}
 1-z=e^{t+i\theta} \ , ~~~~~1-\bar z=e^{t-i\theta} 
\end{align}
and recall that the conformal transformation introduces an
overall factor in the correlator:\footnote{We slightly abuse notation: the coordinate $t$
introduced in \eqref{ztnotation} differs from the time coordinate here. The former simply parametrizes the $(z,\bar z)$
coordinates and does not implement the cylinder-to-plane conformal map. Consequently, the change of variables in \eqref{ztnotation}
does not introduce an additional prefactor in the correlator.}
\begin{align}
\label{Weyl}
 \Phi_0(t,\theta)\rightarrow
\Big((1-z)(1-\bar z)\Big)^{\Delta\over 2}\,\Phi_0(z,\bar z) \ .
\end{align}
Imposing the factorization, $\Phi_0(z,\bar z)=\Phi_0(z)\, \bar\Phi_0(\bar z)$, the equation \eqref{3Ddecouple}  yields the following differential equations: 
\begin{align}
\label{2dBPZlevel2}
 \Big[(1-z)^2\partial_z^2 +6\eta \Big] \Phi_0(z) \Big|_{\Delta= -1} =0 \ , ~~~~~~~~
 \Big[ (1-\bar z)^2\partial_{\bar z}^2 +6\eta \Big] \bar\Phi_0(\bar z)\Big|_{\Delta= -1} =0 \ .
\end{align} 
These are the corresponding level-2 BPZ null-state equations in two-dimensional CFTs at large central charge. In the remainder of our $d=2$ discussion, we will focus on the holomorphic equations.

\subsection{\selectfont\titlefontmedium All-Level Matrix}

The approach described above can be extended to derive higher-level
null-state differential equations. We first express the level-2 equation \eqref{2dBPZlevel2} as
\begin{align}
 \mathscr D_{2}\Phi _0 (z)\Big|_{\Delta= -1} = 0\ , ~~~~~~~  \mathscr D_{2}= (1-z)^2\partial_z^2 +6\eta  \ . 
\end{align}
Rather than setting $\Delta=-1$ in \eqref{3Ddecouple}, we can proceed
to the next order in the near-boundary expansion and express
$\Phi_4$ in terms of $\Phi_0$. We then find that $\Phi_6$ decouples
at $\Delta= -2$. At the subsequent order, the coefficient of
$\Phi_8$ vanishes at $\Delta=-3$, and this decoupling  pattern continues to
higher orders.  Labelling the $d=2$ CFT null-state equations by their level $n\geq 2$, we will organize the differential
equations using an $n\times n$ matrix with a specific pattern of off-diagonal entries.  
This matrix formulation will also provide an efficient way to organize the corresponding equations in higher dimensions.

By extending the holographic computation to higher orders
we find that the bulk coefficient $\Phi_{2n}$ decouples at $\Delta=1-n$, giving rise to the level-$n$ null-state
differential equation. For the next three levels, the gravity computation gives the following (holomorphic) equations
{\setlength{\jot}{2pt}
\begin{align}
&\mathscr D_{3}\Phi _0 (z)|_{\Delta= -2}  = 0 \ , ~~\mathscr D_{3}= (1-z)^3 \partial_z^3 +  \Big(24 (1-z) \partial_z +24 \Big) \eta  \ . \\
&\mathscr D_{4}\Phi _0 (z)|_{\Delta= -3}  = 0 \ , ~~ \mathscr D_{4}=(1-z)^4 \partial_z^4 + \Big(60 (1-z)^2 \partial_z^2 + 120 (1-z) \partial_z  + 108 \Big)  \eta +324 \eta^2 \ . \\
&\mathscr D_{5}\Phi _0 (z)|_{\Delta= -4}  = 0 \ ,   ~~ \mathscr D_{5}= (1-z)^5 \partial_z^5
+ \Big(120 (1-z)^3\partial_z^3 
+360 (1-z)^2 \partial_z^2
+648 (1-z) \partial_z
+ 576 \Big) \eta \nn\\ 
&~~~~~~~~~~~~~~~~~~~~~~~~~~~~~~~~~~~~~~~~~~~~~~~~~ + \Big(2304 (1 - z) \partial_z+ 4608 \Big) \eta^2  \ .
\end{align}}It is straightforward, though increasingly involved, to derive higher-order equations.  The highest power of $\eta$ appearing in
$\mathscr D_n$ increases by one every two levels. 

It is useful to perform the change of variables
\begin{align}
z = 1- e^{-t} \ .
\end{align} 
Denoting the corresponding correlator by $V(t)$, the null-state differential equations at the first four levels become  
{\setlength{\jot}{2pt}
\begin{align}
\mathscr{D}_2 V(t)|_{\Delta= -1} &= 0 \ ,~~~~~  
\mathscr{D}_2=\partial_t^2+\partial_t+6\eta\ . \\
\mathscr{D}_3 V(t)|_{\Delta= -2} &= 0   \ ,~~~~~  
\mathscr{D}_3=\partial_t^3+3\partial_t^2+2\partial_t
+24(1+\partial_t) \eta \ . \\
\mathscr{D}_4 V(t)|_{\Delta= -3} &= 0   \ ,~~~~~ 
\mathscr{D}_4 = \partial_t^4+6\partial_t^3+11\partial_t^2+6\partial_t
+12(9+15\partial_t+5\partial_t^2)\eta +324\eta^2  \ . \\
\mathscr{D}_5 V(t) |_{\Delta= -4}&= 0 \ , ~~~~~ \mathscr{D}_5=\partial_t^5+10\partial_t^4+35\partial_t^3+50\partial_t^2+24\partial_t \nn\\ &~~~~~~~~~~~~~~~~~~~~~~~~~~ +24(24+52\partial_t+30\partial_t^2+5\partial_t^3) \eta +2304 (2+\partial_t) \eta^2  \ . 
\end{align}}Let us first make two simple observations.  First, when the differential
equations are expressed in the basis
$\{\partial_t^k\}_{k=0,1, 2, 3, \ldots}$, their coefficients are
constants, independent of $t$. Second, in the ``free'' case 
$\eta=0$, corresponding to the absence of a black hole in the bulk,
the differential operators exhibit a simple pattern:
\begin{align}
\label{2ddia}
\mathscr{D}_n\Big|_{\eta = 0}  = \partial_t (\partial_t+1)(\partial_t+2)(\partial_t+3) \dots  (\partial_t+n-1) = \prod_{i=1}^{n}(\partial_t+i-1) \ .
\end{align}

In the presence of a black hole, extending the holographic computation to higher orders reveals that the level-$n$ null-state
equation can be expressed as the determinant of an $n\times n$ matrix $M_n$, whose entries for  $i,j=1,\ldots,n$ are given by  
\begin{align}
\label{2dmatricDn}
\mathscr{D}_nV(t)\big|_{\Delta=1-n} = 0\ , ~~~~~~ \mathscr{D}_n= \det {M}_{n}\ , ~~~~~~~  ({M}_n)_{ij} =
\begin{cases}
\d_t+i-1 ~~~~~~ & i=j \\
1 ~~~~~~ & i=j-1 \\
-6 \eta (n-j) j   ~~~~~~ & i=j+1 \\
0 ~~~~~~ &\text{\it otherwise}  
\end{cases}
\end{align} 
The matrix $M_n$ is tridiagonal, with a banded structure in which the diagonal entries increase linearly with $i$, following the pattern in \eqref{2ddia}. Its superdiagonal entries are unity, while its subdiagonal entries have a quadratic profile proportional to the ``coupling" $\eta$ which encodes the effect of the heavy-state background. 
As an illustration, the level-9 matrix is 
{\small\begin{align}
\label{2dn9}
({M}_{9})_{ij} =\left(
\begin{array}{ccccccccc}
 \d_t & 1 & 0 & 0 & 0 & 0 & 0 & 0 & 0 \\
 -48 \eta  & \d_t+1 & 1 & 0 & 0 & 0 & 0 & 0 & 0 \\
 0 & -84 \eta  & \d_t+2 & 1 & 0 & 0 & 0 & 0 & 0 \\
 0 & 0 & -108 \eta  & \d_t+3 & 1 & 0 & 0 & 0 & 0 \\
 0 & 0 & 0 & -120 \eta  & \d_t+4 & 1 & 0 & 0 & 0 \\
 0 & 0 & 0 & 0 & -120 \eta  & \d_t+5 & 1 & 0 & 0 \\
 0 & 0 & 0 & 0 & 0 & -108 \eta  & \d_t+6 & 1 & 0 \\
 0 & 0 & 0 & 0 & 0 & 0 & -84 \eta  & \d_t+7 & 1 \\
 0 & 0 & 0 & 0 & 0 & 0 & 0 & -48 \eta  & \d_t+8 \\
\end{array}
\right)
\end{align}}The level-$n$ differential equation is obtained by taking
the determinant of the corresponding matrix. Certainly, this matrix representation is not unique, since different matrices can have the same determinant. The tridiagonal form adopted here isolates the $\eta$ dependence and makes the banded structure at level $n$ manifest. 
Moreover, the structured off-diagonal entries make its underlying algebraic
structure easier to identify, as we discuss below, and this will provide a useful perspective for generalizing to higher dimensions

\subsection{\selectfont\titlefontmedium Recursion and Product Form}

Based on the $d=2$ matrix \eqref{2dmatricDn}, here we study the characteristic polynomial and derive a recursion relation as well as a factorized product form of the differential equations.

Define the characteristic polynomial
\begin{align}
\label{2dA}
\chi_n(\lambda)=\det\bigl(\lambda{I_n}-{M}_n\bigr) \ , ~~~~ {M}_n=\partial_t{I_n}+{A}_n \ ,
\end{align} 
where $I_n$ is the $n\times n$ identity matrix and $A_n$ is an
$n\times n$ matrix independent of the differential operator
$\d_t$. Since $\partial_t$ commutes with all entries of $A_n$, we can
rearrange the factors in the determinant as usual. 
By writting $\chi_n(\lambda)=\det\bigl((\lambda-\partial_t){I_n}-{A}_n\bigr)$ and letting $\chi_{{A}_n}$ be the characteristic polynomial of ${A}_n$,  
we have  $\chi_n(\lambda)= \chi_{{A}_n}(\lambda-\partial_t)$. 

In addition, denoting the eigenvalues of ${A}_n$ by $\alpha_j$ gives $\chi_n(\lambda)=\prod_{j=1}^n(\lambda-\partial_t-\alpha_j)$.  Therefore,  provided that ${A}_n$ is diagonalizable, the formal roots $\lambda_j=\partial_t+\alpha_j$ are the diagonal entries of ${M}_n$, and  the level-$n$ null-state equation can be written as $\prod^n_{j=1} \lambda_j(\partial_t, \alpha_j) V(t)\Big|_{\Delta=1-n}=0$. 

The characteristic polynomials can be computed explicitly
order by order in $n$. For example,
{\setlength{\jot}{2pt}
\begin{align}
\chi_2(\lambda)&= \left(\lambda-\d_t -1\right) \left(\lambda -\d_t\right)+6 \eta  \ , \\
\chi_3(\lambda)&= \left(\lambda-\d_t -2\right)  \left(\lambda -\d_t -1\right) \left(\lambda - \d_t\right) + 24 \eta  \left(\lambda - \d_t -1\right) \ , \\
\chi_4(\lambda)&= \left(\lambda-\d_t -3\right) \left(\lambda -\d_t-2\right) \left(\lambda  -\d_t-1\right) \left(\lambda - \d_t\right)\nn\\
& ~~~~~~~~~~~~~~ +12 \eta  \left(   5 (\d_t- \lambda )^2+15 (\d_t -\lambda) +9\right) +324 \eta ^2 \ .
\end{align}}

There is a recursive structure which allows a combinatorial representation. The  pattern becomes more apparent by studying the matrix ${A}_n$ for higher values of $n$.   Let ${A}_n^{(j)}$ denote its upper-left $j\times j$ submatrices, $i.e.$, the leading principal submatrices of the matrix ${A}_n$. We define
the corresponding characteristic polynomial as 
\begin{align}
P_j(\lambda)=\det\left(\lambda I_j-A_n^{(j)}\right) \ .
\end{align}  
Thus, $P_{n}= \chi_{A_n}$. 
The tridiagonal structure of ${A}_n$ has a three-term recursion relation:
\begin{align}
\label{2drecursion}
P_j=(\lambda-j+1)P_{j-1} +6\eta(j-1)(n-j+1)P_{j-2} \ ,
~~~ \qquad P_0=1, ~ \quad P_1=\lambda \ .
\end{align} 
For convenience, we introduce 
\begin{align}
a_r(\lambda)=\lambda-r+1 \ , ~~~ b_r (\eta)= 6\eta (n-r)r  \ . 
\end{align} 
The recursion relation becomes $P_j=a_j P_{j-1}+ b_{j-1} P_{j-2}$.
 Iterating this relation gives a finite combinatorial sum:  
\begin{align}
\label{2dPnclosedform}
 P_n (\lambda)=
\sum_{\text{non-overlapping~adjacent ~pairs}}
\left(\prod_{(r,r+1)}b_r(\eta)\right)
\left(\prod_{\text{remaining}}a_\ell (\lambda)\right) \ .
\end{align} 
The sum runs over all collections of non-overlapping adjacent pairs $(r,r+1)$, including the empty collection. 
Each selected pair contributes a factor $b_r$, while each index not belonging to a selected pair contributes a factor $a_\ell$.

Let us illustrate the combinatorial representation \eqref{2dPnclosedform} using $P_4$ as an example. When no pairs are chosen, all four indices contribute their individual factors, giving $a_1a_2a_3a_4$.
If one pair is chosen, there are three possibilities: choosing $r=1$ gives $(b_1a_3a_4)$, choosing $r=2$ gives $(b_2a_1a_4)$, and choosing $r=3$ gives $(b_3a_1a_2)$.   
If two pairs are chosen, the only allowed collection is $(1,2)$ and $(3,4)$, giving $(b_1b_3)$. Notice that the pairs $(1,2)$ and $(2,3)$ cannot be selected simultaneously as they share index $2$.  
Summing all contributions, we obtain $P_4=a_1a_2a_3a_4 +b_1a_3a_4+b_2a_1a_4+b_3a_1a_2+b_1b_3=\left(\lambda^4-6 \lambda^3+11 \lambda^2 -6 \lambda \right)+12 \eta  \left(5 \lambda ^2-15 \lambda +9\right)+324 \eta ^2$.

The eigenvalues of $A_n$ take a simple form. Labeling them by $m$, we write
\begin{align}
\label{2deigen}
\lambda_m=\frac{n-1}{2} (1+\gamma)  -\gamma m \ , ~~~\gamma= \sqrt{1-24 \eta} \ ,  ~~~~ m=0,1,\ldots,n-1 \ .
\end{align} 
To obtain this result, it is more convenient to compute the characteristic polynomials for successive values of $n$ and
identify the pattern in their roots, rather than using expression \eqref{2dPnclosedform}.\footnote{The combinatorial expression
may nevertheless provide a starting point for exploring a graph-theoretic construction of the null-state differential equations, which we leave for future work.} 

As an example, for $n=5$, the diagonalized matrix denoted by $\widetilde M_5$ reads 
\begin{align}
\label{2dM5d}
\small
{\widetilde M}_{5}=\left(
\begin{array}{ccccc}
 \d_t+2  +2\sqrt{1-24 \eta } & 0 & 0 & 0 & 0 \\
 0 & \d_t+2 + \sqrt{1-24 \eta } & 0 & 0 & 0 \\
 0 & 0 & \d_t+2& 0 & 0 \\
 0 & 0 & 0 & \d_t+2-\sqrt{1-24 \eta } & 0 \\
 0 & 0 & 0 & 0 & \d_t+2 -2 \sqrt{1-24 \eta } \\
\end{array}
\right)
\end{align}

In general, the spectrum exhibits the following qualitative behavior:
\begin{itemize}
\item For $\eta < {1\over 24}$, the eigenvalues are real and distinct.
\item For $\eta > {1\over 24}$, the eigenvalues lie on the vertical line ${\rm Re}\,\lambda_m={n-1\over 2}$ in the complex plane and are symmetric about the real axis. 
\item At the critical value $\eta_c={1\over 24}$, all eigenvalues coalesce at $\lambda= {n-1\over 2}$.  
\end{itemize}

It follows that the characteristic polynomial of $M_n$ is 
\begin{align}
\chi_n(\lambda)= P_n(\lambda-\d_t)=\prod _{m=0}^{n-1} \Big[\lambda - \Big(\d_t+\frac{n-1}{2} (1+\gamma)  -\gamma  m \Big)\Big] \ .
\end{align} 
It is useful to introduce a symmetric labeling of the eigenvalues
through
\begin{align}
k=n-1-2m \ , ~~~~~~~~ k=n-1,n-3, n-5, \ldots,-(n-1) 
\end{align} 
where $k$ runs over the symmetric index set. The $d=2$ null-state differential operators  ${ \mathscr{ D}}_n$ take a simple product form  
\begin{align}
\label{2dproductform}
 { \mathscr{ D}}_n = \det M_n= 
\prod_{k=-(n-1)}^{n-1}
\left(
\d_t+\frac{n-1}{2}
+{k\over 2}\sqrt{1-24 \eta}
\right)  \ , ~~~~~   { \mathscr{ D}}_n V(t)\Big|_{\Delta= 1-n} =0 \ .
\end{align} 
This result in $d=2$ can also be derived from the construction of singular vectors in Virasoro representation
theory, without using holography \cite{Bauer:1991ai, DiFrancesco:1997nk}; see also discussions in \cite{Fitzpatrick:2016ive, Chen:2016cms}. Here, starting from AdS$_3$ gravity with a black hole, we re-discover this product formula for the null-state equations at arbitrary levels (in the semiclassical approximation). This provides a consistency check of the holographic computation and the $d=2$ organizing matrix.

\subsection{\selectfont\titlefontmedium Algebraic Structure}

We now show that the $d=2$ matrix \eqref{2dmatricDn} can be expressed as a linear combination of the identity matrix and the generators of $\mathfrak{sl}_2$ in
its $n$-dimensional irreducible representation. We then explore some consequences of this algebraic representation.

Consider three $n\times n$ matrices with entries
\begin{align}
\label{2dJn}
(J^{(n)}_0)_{ij} =  \Big(i-{n+1\over 2}\Big)\delta_{ij} \ , ~~~~~~~  
(J^{(n)}_+)_{ij} = j\big(n-j\big)\delta_{i-1, j} \ ,   ~~~~~~~ 
(J^{(n)}_-)_{ij} =\delta_{i+1, j} \ .
\end{align}
These matrices furnish the $n$-dimensional irreducible representation of the Lie algebra $\mathfrak{sl}_2$, satisfying
the commutation relations
\begin{align}
[{ J}_0, {J}_\pm]=  \pm {J}_\pm \ , ~~~~~~  [{J}_+, {J}_-]=  2{J}_0  \ .
\end{align}
In terms of these generators, the matrix takes the form
\begin{align}
\label{2dMJ}
{M}=\Big(\d_t+\frac{n-1}{2}\Big) { I} +  { J}_0 + {J}_{-} - 6 \eta {J_+} \ .
\end{align}
The differential operator $\d_t$ contributes a uniform shift to the diagonal entries, while the remaining structure is encoded in a linear combination of $\mathfrak{sl}_2$ generators. In particular, the $\eta$-dependent term that encodes the effect of the heavy-state or thermal background, enters
linearly through $J_+$.

The algebraic representation also provides a simple way to derive the spectrum of $M_n$.  Since $X^{(n)} \equiv J_0+J_- -6\eta J_+$ is an element of $\mathfrak{sl}_2$, its spectrum can be determined from the
fundamental representation. By setting $n=2$ in \eqref{2dJn}, the corresponding two-dimensional generators are:\footnote{We adopt the basis \eqref{2dJn} in which the
eigenvalues of $J_0$ are ordered from lowest to highest. Accordingly, the two-dimensional raising and lowering operators have matrix representations that differ from the usual convention.}
\begin{align}
J_0 &= \begin{pmatrix}
-\frac{1}{2} & 0 \\
0 & \frac{1}{2}
\end{pmatrix},
\qquad
J_+ = \begin{pmatrix}
0 & 0 \\
1 & 0
\end{pmatrix},
\qquad
J_- = \begin{pmatrix}
0 & 1 \\
0 & 0
\end{pmatrix}.
\end{align}
The corresponding  ``reduced" matrix is 
\begin{align}
X^{(2)}= J_0+J_--6\eta J_+
=\begin{pmatrix}
-\frac{1}{2} & 1 \\
-6\eta & \frac{1}{2}
\end{pmatrix}
\end{align}
with eigenvalues 
\begin{align}
\lambda_\pm=  \pm {1\over 2} \gamma\ , ~~~~~\gamma= \sqrt{1-24\eta}  \ .
\end{align}
Thus, we see that the critical value $\eta_{\mathrm{c}}=1/24$ can be already determined by the fundamental representation, independent of $n$.   At the critical value,  $X^{(2)}$ is a nonzero nilpotent matrix.

For $\eta\neq\eta_{\mathrm c}$, on the other hand, the reduced matrix $X^{(2)}$ is semisimple and conjugate to $\gamma J_0$ by an element of
$SL(2,\mathbb C)$. To see this, define
\begin{align}
\label{2dS}
S(\gamma)= {1\over \sqrt\gamma}
\begin{pmatrix}
1 & 1 \\[2pt]
\frac{1-\gamma}{2} & \frac{1+\gamma}{2}
\end{pmatrix}  \  , ~~~~ \det S= 1  \ .
\end{align} 
The columns of $S$ are eigenvectors of $X^{(2)}$ with eigenvalues $-\gamma/2$ and $\gamma/2$, respectively.\footnote{Explicitly, for $v_{\pm}=(1,(1\pm\gamma)/2)^{\mathsf T}$, we have $X^{(2)}v_{\pm}=\pm(\gamma/2)v_{\pm}$.} 
For $\gamma\neq0$, the matrix $S$ is invertible, and 
\begin{align}
\label{2dXS}
X^{(2)}=S(\gamma J_0)S^{-1} \ .
\end{align}
Let $R_n(S)$ denote the matrix representing the group element $S\in SL(2,\mathbb C)$ in the $n$-dimensional irreducible representation. The corresponding conjugation in this
representation is
\begin{align}
X^{(n)}
=R_n(S) (\gamma J_0^{(n)})R_n(S)^{-1}  
\end{align}
where $J_0^{(n)}$ is in the $n$-dimensional representation given in \eqref{2dJn}. 
Since conjugation preserves the characteristic polynomial,
the eigenvalues of $X^{(n)}$ are those of
$\gamma J_0^{(n)}$. 
Recalling that $J_0^{(n)}$ has eigenvalues $i-{n+1\over 2}$, the eigenvalues of $X^{(n)}$ are  
\begin{align}
\label{2dresultagain}
\lambda_i\bigl(X^{(n)}\bigr)
&=\left(i-\frac{n+1}{2}\right)\sqrt{1-24\eta}   ~~~~~~~~~ i=1,\ldots,n   \nn\\
 &=\frac{k}{2}\sqrt{1-24\eta}  ~~~~~~~~~~ k=-(n-1),-(n-3),\ldots,n-3,n-1 \ . 
\end{align}
The first expression agrees with \eqref{2deigen}, upon setting $m=n-i$ with the eigenvalues listed in reverse order.  
 The second expression adopts the symmetric index, $k=2i-n-1$. 
Finally, adding the uniform shift
$\d_t+\frac{n-1}{2}$ gives the eigenvalues of $M_n$, and the result 
is consistent with the product formula for the null-state differential 
operators \eqref{2dproductform}.
\\

\noindent {\it Remarks on the critical $\eta$:}  
The conjugation argument above breaks down at the critical value
$\eta_c=1/24$ (or $\gamma=0$), since the conjugating matrix $S(\gamma)$ in \eqref{2dS} becomes singular. Nevertheless, the
spectrum can still be determined by examining the reduced matrix
$X^{(2)}$. At the critical value, we have
\begin{align}
X^{(2)}\big|_{\gamma=0}
=\begin{pmatrix}
-\frac12 & 1 \\[2pt]
-\frac14 & \frac12
\end{pmatrix},
\qquad
\left(X^{(2)}\big|_{\gamma=0}\right)^2=0 \ .
\end{align}
Consider the nonsingular matrix 
\begin{align}
T=\begin{pmatrix}
1 & 0 \\[2pt]
\frac12 & 1
\end{pmatrix},
\qquad
\det T= 1 
\end{align}
which gives
\begin{align}
T^{-1}\big(X^{(2)}\big|_{\gamma=0}\big)T
=\begin{pmatrix}
0 & 1 \\[2pt]
0 & 0
\end{pmatrix} \ .
\end{align}
Therefore, at the critical value, $X^{(2)}$ is conjugate to the lowering generator in our convention and is thus nilpotent.\footnote{Alternatively, one can choose
\begin{align}
\widetilde T=
\begin{pmatrix}
0&i\\
i&\frac{i}{2}
\end{pmatrix} \ , ~~~~ \det \widetilde T= 1  \ . 
\end{align}  
This gives  
\begin{align}
\widetilde T^{-1}\big(X^{(2)}\big|_{\gamma=0}\big) \widetilde T
=\begin{pmatrix}
0 & 0 \\[2pt]
1 & 0
\end{pmatrix} \ .
\end{align}}   
Since $T$ belongs to $SL(2,\mathbb C)$, its conjugation can be carried
over to the $n$-dimensional irreducible representation. Consequently,
$X^{(n)}\big|_{\gamma=0}$ is conjugate to the corresponding
representation of the raising/lowering generator, which is represented by a triangular matrix with vanishing diagonal entries and is therefore nilpotent, $i.e.,$ its $n$-th power vanishes.  
It follows that $X^{(n)}|_{\gamma=0}$ is itself nilpotent, with characteristic polynomial  
\begin{align}
\det(\lambda I-X^{(n)})\big|_{\gamma=0}
=\prod_{j=1}^{n}(\lambda-\lambda_j) \big|_{\gamma=0}
=\lambda^n \ . 
\end{align}
Hence, Eq.~\eqref{2dresultagain} remains valid at the critical value of $\eta$, where all eigenvalues of $X^{(n)}$ simply vanish. 
\\

\noindent {\it Imprint of a larger CFT symmetry:}  From the $d=2$ CFT perspective, the null-state differential equations
are rooted in Virasoro symmetry. This raises the question of whether the differential equations
derived from AdS gravity may point to an infinite-dimensional algebra with a central extension.  
The $\mathfrak{sl}_2$ structure of $M_n$ organizes the
differential equations at arbitrary levels, but it captures only the 
large-$c$, heavy-light, limit of the equations, with finite-$c$
corrections suppressed. Reconstructing the Virasoro algebra from
the matrix therefore appears nontrivial. Nevertheless, the
off-diagonal entries of $M_n$, which are proportional to $\eta$,
retain a dependence on the central charge and may be viewed as a
remnant of the larger (Virasoro) algebraic structure.

\section{\selectfont\titlefontmedium Four-Dimensional Holographic CFT} 
\label{sec:4d}

\subsection{\selectfont\titlefontmedium Review: A  Null-State-Type Equation from AdS$_5$ Black Hole} 

We begin by reviewing the recent holographic derivation in \cite{Huang:2026jbv} of the lowest-level null-state-type differential equation for a light scalar with the degenerate-type scaling dimension $\Delta=-1$.\footnote{The $d=4$ equations discussed in the present work are not derived from Virasoro-type modes $L_m$. We nevertheless use the term ``level'' to distinguish equations
associated with different values of $\Delta$.}

Consider an AdS$_5$-Schwarzschild black hole with a spherical horizon,
described by the metric
 \begin{align}
ds^2 = \big(1 + r^2 f(r)\big)\,dt^2 + \frac{dr^2}{1 + r^2 f(r)} + r^2  d\Omega^2_3 \ , ~~~~~ f(r) = 1- {\mu \over r^4}
\end{align} 
where we set the AdS radius to unity, and $d\Omega_3^2$ denotes the
metric on a unit three-sphere, parametrized by the angular coordinates
$\theta_1,\theta_2,\theta_3$.
We consider a minimally-coupled massive bulk scalar in dual gravity, and the 
bulk-to-boundary propagator obeys the Klein-Gordon equation 
\begin{align}
    \left(-\nabla^2+m^2\right)\Phi=0 \ .
\end{align} 
By rotational symmetry, the angular dependence of the bulk-to-boundary propagator
is determined by the relative angle between the bulk point and the
boundary insertion. We can use a rotation to place the
insertion at a fixed angular position and describe the remaining angular dependence using a single coordinate, $\theta$. 
The bulk equation then depends only on $r$, $t$, and $\theta$.\footnote{This
simplification has been used in previous analyses of the bulk field 
equation, $e.g.$, \cite{Fitzpatrick:2019zqz}. It is also useful when
computing higher-dimensional four-point conformal correlators parametrized only 
by two coordinates $(z,\bar z)$.}
Writing $\Phi=\Phi(r,t,\theta)$, the bulk equation can be written as 
 \begin{align}
 m^2 \Phi  = \Bigg[ \frac{r^2}{r^2 + r^4 - \mu} \partial^2_t + \frac{1}{r^3} \Big( r \partial^2_\theta  + 2 r \cot\theta \partial_\theta 
+ (3r^2 + 5r^4 - \mu) \partial_r + r(r^2 + r^4 - \mu)\,  \partial^2_r \Big) \Bigg] \Phi  \ .
\end{align}   
The corresponding solution in pure AdS takes the same form as \eqref{pureAdS}. 

Including the black hole, following the AdS$_3$ analysis, we consider the near-boundary expansion
 \begin{align}
\label{4dexpansion}
 \Phi (r, t, \theta)= r^{-\Delta} \sum_{i =0, 2, 4, 6, \dots}  {\Phi_{i}(t,\theta)\over r^i} \ .
\end{align} 
The leading-order equation at large $r$ imposes the standard relation
$m^2 = \Delta (\Delta - 4)$. The equation at the next order reads
 \begin{align}
\Big[\partial^2_t +\partial^2_\theta +2\cot\theta \, \partial_\theta + \Delta (\Delta - 2) \Big]\Phi_{0}=  4(1-\Delta ) \Phi_{2} \ .
\end{align} 
We see that the coefficient $\Phi_2$ decouples at $\Delta=1$, for which the
bulk scalar has negative mass squared (but remains above the
Breitenlohner--Freedman bound \cite{breitenlohner1982stability, breitenlohner1982positive}). 
The resulting equation is, however, independent of $\mu$ -- its solution can be readily verified to coincide with the large-$r$ limit of the pure AdS result, \eqref{pureAdS}, evaluated at $\Delta=1$.
We therefore continue to the next order rather than specialize to
this value of $\Delta$. Solving for $\Phi_2$ in terms of $\Phi_0$
and substituting into the equation at the next order yields an 
equation involving only $\Phi_4$ and $\Phi_0$:
\begin{align}
\label{4dphi4}
32\Delta(\Delta-1)\Phi_4 ={}&\Big[\partial_t^4+
\partial_\theta^4+2\partial_t^2\partial_\theta^2
+4\cot\theta \partial_\theta^3+4\cot\theta\partial_t^2\partial_\theta+2\big( \Delta  (\Delta +2)-2\big)\partial_t^2 \nn\\
&\quad  +2(\Delta^2-2)\partial_\theta^2 +4\Delta^2\cot\theta \partial_\theta+\Delta^2(\Delta+2)(\Delta-2)+4\mu\Delta^2 (\Delta-1)
\Big]\Phi_0 \ .
\end{align} 
Unlike in the AdS$_3$ case, $\Phi_4$ does not decouple at any nonzero value of $\Delta$, given the earlier assumption $\Delta\neq 1$. 
We therefore proceed to the next order.
Now using \eqref{4dphi4}  to eliminate $\Phi_4$ in favor of $\Phi_0$,
we obtain, for $\Delta\neq 0,1$, the following equation involving only
$\Phi_6$ and $\Phi_0$:
\begin{align}
384\Delta(\Delta-1)(\Delta+1)\Phi_6
=-\sum_{i+j\leq 6}\alpha_{ij} (\theta)\,
\partial_t^i\partial_\theta^j\Phi_0\,
\end{align}
where the nonzero coefficients $\alpha_{ij} (\theta)$ are given by\footnote{These explicit coefficients were not included in
\cite{Huang:2026jbv}; we provide them here for completeness.} 
{\setlength{\jot}{2.5pt}
\begin{align}
\alpha_{60}&=\alpha_{06}=1,~~~~~~~~  \alpha_{42}=\alpha_{24}=3, \\  
\alpha_{41}&=6\cot\theta, ~~~~~~~~~~ \alpha_{23}=12\cot\theta, ~~~~~~~~ \alpha_{05}=6\cot\theta, \\
\alpha_{40}&=3\Delta^2+18\Delta+4, \\
\alpha_{04}&=3\Delta^2+6\Delta-4, \\
\alpha_{22}&=6\Delta(\Delta+4),\\
\alpha_{03}&=12(\Delta^2+2\Delta+2)\cot\theta,\\
\alpha_{21}&=12(\Delta^2+4\Delta+2)\cot\theta, \\
\alpha_{20}&=3\Delta^4+24\Delta^3+48\Delta^2-24\Delta-32+12\mu\,\Delta^2(\Delta+5), \\
\alpha_{02}&=3\Delta^4+12\Delta^3-24\Delta-32 +4\mu\,\Delta\big(3\Delta^2+7\Delta+8\big), \\
\alpha_{01}&=\Delta\Big(6\Delta^3+24\Delta^2+24\Delta+8\mu\,\big(3\Delta^2+7\Delta+8\big)\Big)\cot\theta,\\
\alpha_{00}&=\Delta^2(\Delta+2)\Big(\Delta^3+4\Delta^2 -4 \Delta -16 +12\mu\,\big(\Delta^2+\Delta-4\big)\Big) \ .
\end{align}}These coefficients are independent of $t$. We see that $\Phi_6$ decouples at $\Delta= -1$, reminiscent of the
decoupling of the corresponding bulk coefficient in the AdS$_3$ case. 

Restricting the $d=4$ scalar correlator to a two-dimensional plane, as usual, we adopt $z,\bar z$ coordinates through the map \eqref{map}, taking into account the Weyl factor \eqref{Weyl}.   
In addition, to simplify the expressions in the $d=4$ case, we find it helpful to write\footnote{The Weyl factor \eqref{Weyl} depends on $\Delta$, and we will adopt this $\Delta$-independent prefactor to simplify the $d=4$ differential equations at different values of $\Delta$.}
 \begin{align}
\label{m1QF}
\Phi_{0}(z,\zb) = \frac{1-\bar z}{z-\bar z}\, Q  (z,\zb)   \ .
\end{align} 
The CFT differential equation, here for $\Delta = -1$, can be written as \cite{Huang:2026jbv}  
 \begin{align}
\label{4dEq1}
\Big[ \, {\cal D}^{(3)}  {\cal \overline D}^{(3)}  - 3 {\cal D}^{(3)}  {\cal \overline D}^{(2)}  
+ \mu \big({\cal D}^{(2)}   +  {\cal \overline D}^{(2)} + {\cal D}  {\cal \overline D} + {\cal D} \big)\, \Big] \, Q(z,\zb)  \Big|_{\Delta = -1} = 0  
\end{align} 
where we introduce the weighted differential operators ($\partial= \partial_z$, $\bar \partial=\partial_{\zb}$):
 \begin{align}
{\cal D}^{(n)}\equiv (1-z)^n \partial^n  \ , ~~~~~ {\cal \overline D}^{(n)}\equiv (1-\bar z)^n \bar\partial^n \ .
\end{align} 

\noindent {\it Near-lightcone differential equations:} It was shown in \cite{Huang:2026jbv} that, in the lightcone limit,
the differential equation \eqref{4dEq1} reduces to\footnote{As usual, $\tau$ and $J$ denote the twist and spin, respectively,
of the exchanged operators. The limit $\bar z\to 0$, which is
 formal in Euclidean signature, corresponds to the
lightcone limit in Lorentzian signature, where $z$ and $\bar z$
become independent real variables. In this limit, the contributions
from higher-twist operators are suppressed, following from the near-lightcone structure of the $d=4$ global conformal block \cite{Dolan:2000ut}:
\begin{align}
\label{lcblock}
(z\bar z)^{\frac{\tau}{2}}
\left(-\frac{z}{2}\right)^J
{}_2F_1\left(
\frac{\tau}{2}+J,\frac{\tau}{2}+J,
\tau+2J;z
\right)
+{\cal O}\left(\bar z^{\frac{\tau}{2}+1}\right) \ .
\end{align}  
Since the conformal block carries an overall factor of
 $\bar z^{\tau/2}$, we isolate the minimal-twist contribution by
taking $\bar z\to 0$ while holding $v=\mu\bar z$ fixed.
After isolating the minimal-twist contribution, the explicit $\mu$ dependence
can be restored in the differential equation.}
\begin{align}
\label{4dQeq1}
\bar\partial^2 \Big(\mu+(1-z)^3\partial^3\bar\partial\Big)
Q^{\tau{\rm min}}(z,\zb)\Big|_{\Delta = -1}=0 \ .
\end{align} 
An equivalent form is 
\begin{align}
\label{4dF0before}
\Bigg[(1+\bar z\bar\partial)\left(\partial^3+\frac{6}{z}\partial^2
+\frac{6}{z^2}\partial\right)+\frac{\mu\bar z}{(1-z)^3}\Bigg] \,
F^{\tau_{\rm min}}(z, \zb)\Big|_{\Delta = -1}=\alpha(z)+\beta(z)\bar z 
\end{align}
where we introduce $F^{\tau_{\rm min}}(z, \zb)$ by factoring out $(z\bar z)^\Delta$ from $\Phi^{\tau_{\rm min}}_{0}$, for general $\Delta$. More explicitly, 
 \begin{align}
F^{\tau_{\rm min}} \Big|_{\Delta = -1} &= (z\,\bar z)^{\Delta} \, \Phi^{\tau_{\rm min}}\Big|_{\Delta = -1} = \frac{1}{z^2\,\bar z} \,  Q^{\tau_{\rm min}} \Big|_{\Delta = -1} \ . 
\end{align} 
This relation is consistent with the $\Delta$-independent relation \eqref{m1QF} in a small-$\bar z$ limit. The functions $\alpha(z)$ and $\beta(z)$ arise as integration
constants after removing the overall derivatives $\bar\partial^2$ in \eqref{4dQeq1}.

To fix $\alpha(z)$ and $\beta(z)$, we consider the near-lightcone expansion
\begin{align}
\label{lightconeF}
F^{\tau_{\rm min}}(z,\bar z) =\sum_{k=0}^{\infty}\, g_k(z)(\mu\bar z)^k 
\end{align}
where $g_0=1$ corresponds to the identity contribution and
\begin{align}
\label{g14d}
g_1(z)=\frac{\Delta}{120}\, z^3{}_2F_1(3,3;6;z) 
\end{align}
is the small-$\bar z$ limit of the single-stress tensor (global) block. The consistency with $g_0$ and $g_0$ fixes $\alpha(z)=\beta(z)=0$. The resulting equation governs all minimal-twist multi-stress tensor exchanges, with the sum over spins encoded in the functions $g_k(z)$ for $k>1$.

The $d=4$ near-lightcone equation, \eqref{4dF0before} with
$\alpha(z)=\beta(z)=0$, $i.e.$,
 \begin{align}
\label{4dFeq1}
\Bigg[(1+\bar z\bar\partial)\left(\partial^3+\frac{6}{z}\partial^2
+\frac{6}{z^2}\partial\right)+\frac{\mu\bar z}{(1-z)^3}\Bigg] \,
F^{\tau_{\rm min}}(z, \zb)\Big|_{\Delta = -1}=0\ ,
\end{align}
was first conjectured in \cite{Huang:2023ikg}, motivated by an analogy with the level-2 BPZ null-state equation and by crossing symmetry of the scalar four-point correlator. 
The equation has passed several nontrivial checks. In particular, it reproduces the resummed contributions from minimal-twist double- and triple-stress tensor exchanges, previously obtained through
independent methods, including holographic calculations \cite{Kulaxizi:2019tkd} and the lightcone bootstrap technique \cite{Karlsson:2019dbd, Karlsson:2020ghx}.

Let us briefly review this consistency check by explicitly solving the differential
equation order by order in the near-lightcone expansion. 
It is convenient to first absorb $\mu$ into
$\bar z$ and write
\begin{align}
F^{\tau_{\rm min}}(z,\bar z)
=1+g_1(z)\bar z+g_2(z)\bar z^2+g_3(z)\bar z^3
+{\cal O}(\bar z^4)\,.
\end{align} Substituting this expansion into \eqref{4dFeq1} gives a
recursive set of ordinary differential equations:
{\setlength{\jot}{2pt}
\begin{align}
\label{4dm1pereq}
0=& \, \Big(2 z^2 \, g_1^{'''}(z)+12 z \, g_1''(z)+12 \, g_1'(z)+\frac{z^2}{(1-z)^3}\Big)  \, \z \nn\\
&~~ + \Big(3 z^2 \, g_2^{'''}(z)+18 z \, g_2''(z) +18 \, g_2'(z) +\frac{z^2}{(1-z)^3} \, g_1(z) \Big) \, \z^2  \nn\\
&~~~~~~ +\Big(4 z^2 \, g_3^{'''}(z)+ 24  z \, g_3''(z)+24 \, g_3'(z) +\frac{z^2 }{(1-z)^3} \,g_2(z) \Big) \,\z^3 + {\cal O}\left(\z^4\right) \ . 
\end{align}}At first order, the solution is
\begin{align}
g_1(z)= {1\over 4 z^2 } \Big( c_1 + c_2 z + c_3 z^2 +\big(6+(z-6) z\big) \ln (1-z)\Big)  \ .
\end{align} 
The integration constants $(c_1, c_2, c_3)$ can be
fixed by examining the small-$z$ behavior: imposing
$g_1(0)=0$ gives $(c_1, c_2, c_3) = (0, 6, -3)$.
The result  
\begin{align}
g_1(z)={1\over 4 z^2 }  \Big[\big(6+(z-6) z\big) \ln (1-z)-3 (z-2) z\Big]  
\end{align}  reproduces the single-stress tensor contribution in the small-$\zb$ limit, given in \eqref{g14d}, at $\Delta= -1$. 
The same procedure applies at higher orders. Solving the
equations for $g_2$ and $g_3$ gives
{{\setlength{\jot}{2pt}
\begin{align}
g_2&= {1\over 48 z^2} \Big[\big( (z+6)z-6\big) \ln ^2(1-z)-9 (z-2) z \ln (1-z) + 24 z^2\Big] \ , \\
g_3&= {1\over 1152 z^2} \Big[\big( (z-18) z+18\big) \ln^3(1-z) -18 (z-2) z \ln^2 (1-z)  \nn\\
&~~~~~~~~~~~~~~~+ \big(3 (41 z-210) z +630 \big) \ln (1-z) -315 (z-2) z \Big] \ . 
\end{align}}The expression for $g_2$ agrees with the resummed
double-stress tensor contribution to the correlator obtained first 
in \cite{Kulaxizi:2019tkd} using multi-stress tensor OPE
coefficients computed from gravity.  The expression
for $g_3$ agrees with the results of
\cite{Karlsson:2019dbd, Karlsson:2020ghx}, obtained using
the analytic lightcone bootstrap \cite{Fitzpatrick:2012yx, Komargodski:2012ek}. 

Two additional near-lightcone differential equations, with $\Delta= -2, -3$,  in $d=4$
holographic CFTs, were also put forward in \cite{Huang:2023ikg}:
{\setlength{\jot}{2pt}
\begin{align}
\label{4dlightconeDelta2}
& \Bigg[\Big( 2+\z \bar \partial \Big) \left( \partial^4 + {12\over z}  \partial^3 +{36\over z^2}  \partial^2 +{24 \over z^3}  \partial \right) 
+\frac{6 \mu \z }{(1-z)^3}\Big( \partial+ \frac{3 (2-z)}{2  z(1-z) } \Big) \Bigg]\, F(z,\z)\Big|_{\Delta = -2} = 0  \ , \\
\label{4dlightconeDelta3}
& \Bigg[ \Big(3+\z \bar\partial \Big) \left( \partial^5 + \frac{20}{z} \partial^4 +\frac{120}{z^2} \partial^3  +\frac{240}{z^3} \partial^2 +
\frac{120}{z^4} \partial  \right)  \nn\\
&~~~~~~~~~~~~~~~~~~~~~~~~~ + {21 \mu \z \over (1-z)^3} \Big( \partial^2 +\frac{8-5 z }{(1-z) z}  \partial + \frac{24 z^2-84 z+84 }{7 (1-z)^2 z^2} \Big) \Bigg] \, F(z,\z)\Big|_{\Delta = -3} = 0 \ .
\end{align}}These equations were subsequently derived from gravity in
\cite{Huang:2026jbv}, which also obtained an additional
equation for $\Delta= -4$. As in the AdS$_3$ case, the
near-boundary expansion of the bulk equation truncates
at these degenerate-type values of $\Delta$, yielding 
the dual CFT differential equations.  Discussions of the solutions to these Fuchsian-type differential
equations can be found in \cite{Huang:2024wbq}.

\subsection{\selectfont\titlefontmedium All-Level Matrix}

Obtaining CFT differential equations at larger $|\Delta|$ requires expanding the bulk field equation to higher orders, making the calculations increasingly cumbersome. Here, we construct a matrix that generalizes its $d=2$
counterpart and generates null-state-type equations for holographic $d=4$ CFTs at arbitrary degenerate-type values of $\Delta$ at large central charge. This approach applies to general kinematics and does not require taking the lightcone limit.

By extending the AdS$_5$ calculation to higher orders in the near-boundary expansion, we observe a pattern in which $\Phi_{2n}$
decouples at $\Delta=2-n$, for integers $n\geq 3$. This is analogous to the AdS$_3$ case, where $\Phi_{2n}$
instead decouples at $\Delta=1-n$, for integers $n\geq 2$. The two cases therefore differ by a shift in $\Delta$. In this convention, the first $d=4$
null-state-type equation, given in \eqref{4dEq1}, corresponds to ``level"  $n=3$.  

First recall the correlator relation \eqref{m1QF}, which relates
$\Phi_0(z,\bar z)$ to $Q(z,\bar z)$. Motivated by the $d=2$ analysis,
we introduce the coordinates $t$ and $\bar t$ through
\begin{align}
\label{zt}
z = 1-e^{-t}\ , \qquad \bar z = 1-e^{-\bar t}
\end{align}
and express $Q(z,\bar z)$ in terms of $V(t,\bar t)$.
The first three differential equations can be written as 
{\setlength{\jot}{2pt}
\begin{align}
\label{4dV1}
\mathscr{D}_3 V(t,\bar t)\big|_{\Delta=-1}=0 \ , ~~~
\mathscr{D}_3
&=\partial_t(\partial_t+1)(\partial_t+2)
\,(\partial_{\bar t}-1)\partial_{\bar t}(\partial_{\bar t}+1)
 +\mu\,\Big(
\partial^2_t+\partial_t\partial_{\bar t}
+\partial^2_{\bar t}+2 \partial_t+\partial_{\bar t}
\Big) \ . \\
\label{4dV2}
\mathscr{D}_4 V(t,\bar t)\big|_{\Delta=-2}=0 \ , ~~~
\mathscr D_4
&=
\partial_t(\partial_t+1)(\partial_t+2)(\partial_t+3)
(\partial_{\bar t}-1)\partial_{\bar t}
(\partial_{\bar t}+1)(\partial_{\bar t}+2)
\nn\\
&~~~ +3\mu\,\Big[
\partial_t^3(2\partial_{\bar t}+1)
+\partial_t^2\big(
2\partial_{\bar t}^{\,2}
+11\partial_{\bar t}+5
\big)
\nn\\
&\qquad
+\partial_t\big(
2\partial_{\bar t}^{\,3}
+9\partial_{\bar t}^{\,2}
+16\partial_{\bar t}+6
\big)
+3\partial_{\bar t}(\partial_{\bar t}+1)(\partial_{\bar t}+2)
\Big]
+9\mu^2 \ . \\
\nn\\
\label{4dV3}
\mathscr D_5 V(t,\bar t)\big|_{\Delta=-3}=0 \ , ~~~
\mathscr D_5
&=
\partial_t(\partial_t+1)(\partial_t+2)
(\partial_t+3)(\partial_t+4)
 (\partial_{\bar t}-1)
\partial_{\bar t}
(\partial_{\bar t}+1)(\partial_{\bar t}+2)
(\partial_{\bar t}+3)
\nn\\
&~~~~~ +3\mu\,\Big[\partial_t^4
\big(7\partial_{\bar t}^{\,2}
+14\partial_{\bar t}+3
\big)
+\partial_t^3\big(
7\partial_{\bar t}^{\,3}
+77\partial_{\bar t}^{\,2}
+129\partial_{\bar t}+27
\big)\nn\\
&\qquad~~~~~~~~
+\partial_t^2\big(
7\partial_{\bar t}^{\,4}
+70\partial_{\bar t}^{\,3}
+297\partial_{\bar t}^{\,2}
+388\partial_{\bar t}+78
\big)
\nn\\
&\qquad~~~~~~~~
+\partial_t\big(
28\partial_{\bar t}^{\,4}
+192\partial_{\bar t}^{\,3}
+476\partial_{\bar t}^{\,2}
+432\partial_{\bar t}+72
\big)
\nn\\
&\qquad~~~~~~~~
+24\partial_{\bar t}(\partial_{\bar t}+1)
(\partial_{\bar t}+2)(\partial_{\bar t}+3)
\Big]
\nn\\
&~~~~~ +36\mu^2\,\Big[
\partial_t^2+5\partial_t\partial_{\bar t}
+\partial_{\bar t}^{\,2}
+9\partial_t+12\partial_{\bar t}+18
\Big].
\end{align}}The equation \eqref{4dV1} provides an alternative expression for
\eqref{4dEq1}, while \eqref{4dV2} and \eqref{4dV3} generalize
\eqref{4dlightconeDelta2} and \eqref{4dlightconeDelta3},
respectively, to general kinematics without taking the lightcone limit.
 
In the absence of a black hole in dual gravity, the expressions above exhibit a simple pattern:
{\setlength{\jot}{2pt}
\begin{align}
 \mathscr{D}_n \, \big|_{\mu=0} &= \partial_t(\partial_t+1)\cdots(\partial_t+n-1)\, 
(\partial_{\bar t}-1)\partial_{\bar t}
(\partial_{\bar t}+1)\cdots(\partial_{\bar t}+n-2) \nn\\
&= \prod_{i=1}^{n} (\partial_t+i-1)(\partial_{\bar t}+i-2) \ .
\end{align}}When a black hole is included in dual gravity, the differential equations acquire additional terms depending on $\mu$. 
By deriving the equations at higher levels, whose explicit expressions we omit here, we find that the $d=4$ CFT null-state-type equations can be
organized into the following $n\times n$ matrix
($i,j=1,\ldots,n$):\footnote{Generalizing the $d=2$ matrix in \eqref{2dmatricDn}, one can 
make an $n\times n$ matrix ansatz for the $d=4$ case by allowing
additional off-diagonal entries, requiring its determinant to
reproduce the holographic differential equations at several levels,
and then test whether the resulting matrix correctly predicts the
equations at higher levels.}
\begin{align}
\label{4dmatricDn}
\mathscr{D}_n V(t,\bar t)\big|_{\Delta=2-n}=0\ , ~~~~~ \mathscr{D}_n= \det {M}^{(n)} \ , ~~~~~ (M^{(n)})_{ij}
=\begin{cases}
(\d_t+i-1)(\partial_{\bar t}+i-2) ~~~~~& i=j \\[3pt]
i \, \big(\d_t+\partial_{\bar t}+2(i-1)\big) ~~~~~& i=j-1\\[3pt]
i \, (i+1) & i=j-2\\[3pt]
\dfrac{\mu}{4}(n-i+1) (n-i+2)  ~~~~~& i=j+2\\[3pt]
0 ~~~~~ & \it{otherwise} 
\end{cases}\nn\\ 
\end{align} 
To illustrate the matrix structure, we give the example with $n=6$,
corresponding to $\Delta=-4$:
{\small\begin{align}
\label{4dM6}
{M}^{(6)}= \left(
\begin{array}{cccccc}
 \d_t \left(\partial_{\bar t}-1\right) & \d_t+\partial_{\bar t} & 2 & 0 & 0 & 0 \\
 0 & \left(\d_t+1\right) \partial_{\bar t} & 2 \left(\d_t+\partial_{\bar t}+2\right) & 6 & 0 & 0 \\
 5 \mu  & 0 & \left(\d_t+2\right) \left(\partial_{\bar t}+1\right) & 3 \left(\d_t+\partial_{\bar t}+4\right) & 12 & 0 \\
 0 & 3 \mu  & 0 & \left(\d_t+3\right) \left(\partial_{\bar t}+2\right) & 4 \left(\d_t+\partial_{\bar t}+6\right) & 20 \\
 0 & 0 & \frac{3 \mu }{2} & 0 & \left(\d_t+4\right) \left(\partial_{\bar t}+3\right) & 5 \left(\d_t+\partial_{\bar t}+8\right) \\
 0 & 0 & 0 & \frac{\mu }{2} & 0 & \left(\d_t+5\right) \left(\partial_{\bar t}+4\right) \\
\end{array}
\right)\nn\\
\end{align}}}

Let us compare the $d=2$ and $d=4$ matrices, given in \eqref{2dmatricDn} and \eqref{4dmatricDn}, respectively:
\begin{enumerate}
\item {\it Matrix bandwidth:} The $d=4$ matrix has nonzero entries
on the main diagonal, the first and second superdiagonals, and the
second subdiagonal.  The $d=2$ matrix is tridiagonal,
with nonzero entries only on the main diagonal and the first super- and subdiagonals.
\item {\it Diagonal entries:} The diagonal entries of the $d=4$
matrix are products of operators involving both $\partial_t$ and
$\partial_{\bar t}$, whereas those of the $d=2$ matrix involve
only $\partial_t$.
\item {\it Off-diagonal structure:} In the $d=4$ matrix, the first
superdiagonal entries depend on both derivatives, the second
superdiagonal entries are numerical constants, and the second
subdiagonal entries are proportional to $\mu$ that depends on the central charge. In the $d=2$ matrix, the first superdiagonal entries are unity, while the first subdiagonal entries are proportional to the parameter $\eta$.
\end{enumerate}

The $d=4$ matrix form \eqref{4dmatricDn} directly generalizes its $d=2$ counterpart \eqref{2dmatricDn} and it will allows us to identify an algebraic form. Unlike the $d=2$ matrix, the $d=4$ matrix does not appear to admit a comparably simple diagonalized form, such as the one given in \eqref{2dM5d}, as we will discuss further below. Nevertheless, the patterned off-diagonal structure of the $d=4$ matrix provides an efficient way to organize the differential equations at an arbitrary level, without requiring a simple diagonalized form.

\subsection{\selectfont\titlefontmedium Recursion and Product Form} 

A recursion relation can be derived from the $d=4$ matrix \eqref{4dmatricDn}. Recall that, in the $d=2$ case we define $A_n$ in \eqref{2dA} by removing the diagonal operator $\d_t$ from the matrix. In $d=4$, however, the diagonal entries involve products of $\d_t$ and $\partial_{\bar t}$, so it is more direct to work with the full matrix. Furthermore, the $d=4$ matrix has both a second superdiagonal and a second subdiagonal. Expanding its determinant along the last row will introduce a determinant of a submatrix with a row and a column deleted, $i.e.$, a minor, rather than yielding a recursion involving only smaller principal determinants.

More precisely, let $M_n^{(j)}$ denote the upper-left $j\times j$ principal submatrix of $M_n$. We define
\begin{align}
P_j=\det M_n^{(j)},\qquad P_0=1  
\end{align}
and introduce an auxiliary minor 
\begin{align}
Q_j=\det\widetilde M_n^{(j-1)}  
\end{align}
where $\widetilde M_n^{(j-1)}$  is a $(j-1)\times(j-1)$ matrix, with $j>2$, obtained by deleting rows $j, j+1, j+2, \ldots,n$ and columns $j-2,j+1,j+2, j+3, \ldots,n$ from the full matrix $M_n$.  
Since the last row of the $d=4$ matrix \eqref{4dmatricDn}
contains only two nonzero entries, namely the diagonal entry
and the second-subdiagonal entry, expanding $P_j$ along the 
row relates it to the smaller principal determinant $P_{j-1}$ and the auxiliary minor $Q_j$. The recursion relation is 
\begin{align}
P_j={}&(\d_t+j-1)(\partial_{\bar t}+j-2)P_{j-1}+\frac{\mu}{4}(n-j+1) (n-j+2) Q_j \ ,
\qquad n\geq  j\geq 3 \ .
\end{align}  
This differs from the three-term  recursion  obtained
in the $d=2$ case, given in \eqref{2drecursion}, which follows
from the tridiagonal structure of the $d=2$ matrix. It would be interesting to express $P_n$ in terms of a combinatorial
structure analogous to the $d=2$ expression \eqref{2dPnclosedform}, but here we will not pursue such an expression from the $d=4$ recursion relation. 
 
We now turn to the question of whether the $d=4$ null-state-type equations admit a product form analogous to the $d=2$ expression 
\eqref{2dproductform}. To explore the related structure, let us take a detour and represent the $d=4$ matrix in terms of a differential
operator. Let 
\begin{align}
\label{4ddif}
\mathcal D_n
={}&\left[(1+u)^2+\frac{\mu}{4}u^4\right]\partial_u^2
+\left[(\d_t+\partial_{\bar t})(1+u)
+\mu\left(1-\frac n2\right)u^3\right]\partial_u
\nn\\
&+\d_t(\partial_{\bar t}-1)
+\frac{\mu}{4}(n-1)(n-2)u^2  \ . 
\end{align}
We also introduce the monomial basis
\begin{align}
e_k(u)=u^k,\qquad k=0,\ldots,n-1 \ .
\end{align}
For a fixed $k$, the action of $\mathcal D_n$ on $f_k(t,\bar t) \, e_k(u)$, where $f_k(t,\bar t)$ is a differentiable function, reads 
\begin{align}
\label{CaseCMonomialAction}
\mathcal D_n\big(f_k(t,\bar t)u^k\big)
={}&\big(k+\d_t\big)\big(k+\partial_{\bar t}-1\big)\,  f_k(t,\bar t) u^k
\nn\\
&+k\big(\d_t+\partial_{\bar t}+2k-2\big)\,  f_k(t,\bar t) u^{k-1}
+k\big(k-1\big)f_k(t,\bar t) \, u^{k-2}
\nn\\
&+\frac{\mu}{4}\, \big(k-n+1\big)\big(k-n+2\big)f_k(t,\bar t)\,  u^{k+2}.
\end{align}
Consider an expansion in this basis
\begin{align}
\Psi(u,t,\bar t)=\sum_{j=0}^{n-1}f_j(t,\bar t)e_j(u) \ .
\end{align}
Applying \eqref{CaseCMonomialAction} to this expansion and collecting the coefficients of the basis monomials, we obtain
\begin{align}
\label{CaseCMatrixRepresentation}
\mathcal D_n\Psi
=\sum_{i,j=0}^{n-1}
e_i(u) \big[(M_n)_{i+1,j+1}f_j(t,\bar t)\big] 
\end{align}
where the differential-operator entries $(M_n)_{ij}$ are
given in \eqref{4dmatricDn}. Thus, the action of
$\mathcal D_n$ on the monomial basis encodes the matrix
$M_n$, with its entries acting on the corresponding
coefficient functions.

The space of polynomials in $u$ of degree at most $n-1$, with differentiable
coefficient functions of $t$ and $\bar t$, is invariant under $\mathcal D_n$. Indeed, the coefficient of $u^{k+2}$ in \eqref{CaseCMonomialAction}  vanishes for $k=n-1, n-2$.   For $0\leq k\leq n-3$, the resulting power satisfies $k+2\leq n-1$ and therefore remains within the polynomial space spanned by $\{1,u,\ldots,u^{n-1}\}$.  Since the derivatives $\partial_t$ and $\partial_{\bar t}$ act only on the coefficient functions, they do not change the degree in $u$. 
Thus, $\mathcal D_n$ maps this polynomial space into itself, allowing us to represent the $d=4$ matrix $M_n$ through the action of the differential operator on a finite-dimensional polynomial space.

It is useful to introduce differential operators
\begin{align}
\widehat{\alpha}=\frac{\d_t-\partial_{\bar t}+1}{2} \ ,
\qquad
\widehat{s}=\frac{\d_t+\partial_{\bar t}-1}{2} \ . 
\end{align}
They satisfy $[\widehat{\alpha},\widehat{s}]=0$, $\d_t=\widehat{s}+ \widehat{\alpha}$ and 
$\d_t(\partial_{\bar t}-1)=\widehat{s}^{\,2}-\widehat{\alpha}^{\,2}$. 
For a given $k$, the action of $\mathcal D_n$ on $f_k(t,\bar t)u^k$ can be rewritten as
\begin{align}
\label{CaseCMonomialActionShifted}
\mathcal D_n\big(f_k(t,\bar t)u^k\big)
={}&\big[(\widehat{s}+k)^2-\widehat{\alpha}^{\,2}\big]
f_k(t,\bar t)\,u^k
\nn\\
&+k(2\widehat{s}+2k-1)f_k(t,\bar t)\,u^{k-1}+k(k-1)f_k(t,\bar t)\,u^{k-2}
\nn\\
&+\frac{\mu}{4}(k-n+1)(k-n+2)
f_k(t,\bar t)\,u^{k+2}
\end{align}
where $\widehat{s}$ and $\widehat{\alpha}$ act only on $f_k(t,\bar t)$.
Since $\widehat{\alpha}$ enters only through the diagonal terms, we define $B_n(\widehat{s})$ by separating the
 $-\widehat{\alpha}^{\,2}$ contribution from the diagonal of $M_n$, and write 
\begin{align}
\label{CaseCMatrixRelation}
M_n=B_n(\widehat{s})-\widehat{\alpha}^{\,2}I_n \ . 
\end{align}
Again, $I_n$ is the $n\times n$ identity matrix, and here $B_n(\widehat{s})$ depends only on $\widehat{s}$ -- its
entries are differential operators acting on the coefficient functions $f_j(t,\bar t)$, with the diagonal entries given by 
$(\widehat{s}+k)^2$, $k=0,\ldots,n-1$. The polynomial space remains invariant under this slightly modified operator $B_n$ since the
coefficients of the potentially out-of-range terms $u^{k+2}$ still vanish for $k=n-1, n-2$.

As an example, consider the case $n=3$. Applying \eqref{CaseCMonomialActionShifted} to
\begin{align}
\Psi=f_0(t,\bar t)+f_1(t,\bar t)u+f_2(t,\bar t)u^2
\end{align}
and collecting terms with equal powers of $u$, we get 
\begin{align}
\mathcal D_3\Psi
={}&\big[(\widehat{s}^{,2}-\widehat{\alpha}^{2})f_0
+(2\widehat{s}+1)f_1+2f_2\big]\nn\\
&+\big[\big((\widehat{s}+1)^2-\widehat{\alpha}^{2}\big)f_1
+(4\widehat{s}+6)f_2\big] \, u
+\big[\frac{\mu}{2}f_0
+\big((\widehat{s}+2)^2-\widehat{\alpha}^{2}\big)f_2\big] \, u^2 \ .
\label{CaseCExampleN3}
\end{align}
Comparing this expression with \eqref{CaseCMatrixRepresentation}, and using $M_3=B_3(\widehat{s})-\widehat{\alpha}^{\, 2} I_3$, we see 
\begin{align}
B_3=
\begin{pmatrix}
\widehat{s}^2 & 2\widehat{s}+1 & 2\\
0 & (\widehat{s}+1)^2 & 4\widehat{s}+6\\
{\mu\over 2} & 0 & (\widehat{s}+2)^2
\end{pmatrix}
\end{align}
which gives the determinant 
\begin{align}
\det M_3
=(\widehat{s}^{\,2}-\widehat{\alpha}^{\,2})
\big((\widehat{s}+1)^2-\widehat{\alpha}^{\,2}\big)
\big((\widehat{s}+2)^2-\widehat{\alpha}^{\,2}\big)
+\mu\Big[
(2\widehat{s}+1)(2\widehat{s}+3)
-\big((\widehat{s}+1)^2-\widehat{\alpha}^{\,2}\big)
\Big] \ .
\label{CaseCDeterminantN3}
\end{align}
This reproduces the differential equation given in \eqref{4dV1}, after recalling that $\mathscr{D}_3 V(t, \bar t)=0$ with $\mathscr{D}_3= [\det M_3 (\d_t, \partial_{\bar t})]$. The example illustrates how to compute the determinant using the operator defined in \eqref{4ddif}.  

The $n=3$ example also shows that the eigenvalues of $B_3$ are roots of a nontrivial polynomial with coefficients formally depending on $s$ and $\mu$. The eigenvalues are algebraic functions of these parameters rather than simple polynomial expressions. More generally, let $\lambda_j(s)$, $j=1,\ldots,n$, denote the eigenvalues of $B_n$, regarded as functions of the scalar parameter $s$. We then formally replace $s$ by the differential operator $\widehat{s}$ to write 
\begin{align}
\mathscr{D}_n
=\det M_n
=\prod_{j=1}^{n}
\left[-\Big(\frac{\d_t-\dtb+1}{2}\Big)^{2}+\lambda_j(\widehat{s})\right],
\qquad  \qquad 
\mathscr{D}_n V(t,\bar t)\big|_{\Delta=2-n}=0 \ .
\label{4dproductform}
\end{align}
This $d=4$ expression is the analogue of the $d=2$ product form \eqref{2dproductform} for the null-state differential equations. In both cases, the determinant is expressed as a product over the eigenvalues of a finite-dimensional matrix. In $d=2$, the eigenvalues have a simple closed-form expression, leading to an explicit product formula.  The $d=4$ matrix can still be diagonalized, with eigenvalues instead given by roots of algebraic equations.  We here express the $d=4$ product form through the eigenvalues of $B_n$, which can be determined from the differential operator \eqref{4ddif}.

\subsection{\selectfont\titlefontmedium Algebraic Structure} 

We now identify the algebraic structure underlying the $d=4$ matrix governing the null-state-type differential equations. We consider the generators
\begin{align}
\label{4dJn}
(J_0)_{ij}=\left(i-\frac{n+1}{2}\right)\delta_{ij} \ , ~~~~~
(J_+)_{ij}=(j-n)\delta_{i,j+1} \ , ~~~~~
 (J_-)_{ij}= -i\,\delta_{i,j-1} 
\end{align}  with $i,j=1,\ldots,n$. 
These generators realize an $n$-dimensional irreducible representation of the Lie algebra $\mathfrak{sl}_2$: 
\begin{align}
[J_0,J_\pm]=\pm J_\pm \ , ~~~~~~~ [J_+,J_-]= 2J_0 \ .
\end{align} 
The $d=4$ matrix, given in \eqref{4dmatricDn}, can be expressed as
\begin{align}
\label{4dMJ}
M_n=
\left[\left(\d_t+\frac{1}{2}\right)I_n+\A_n\right]&
\left[\left(\partial_{\bar t}-\frac{1}{2}\right)I_n+\A_n\right]
+\frac{\mu}{4}J_+^2 
\end{align} 
where $I_n$ is the $n\times n$ identity matrix, and we define
\begin{align}
\A_n=\frac{n-2}{2}I_n+J_0-J_- \ . 
\end{align} 
The expression \eqref{4dMJ} generalizes the algebraic form of the $d=2$ matrix \eqref{2dMJ}.

Several comments are in order:

\begin{enumerate} 

\item 
In both $d=2$ and $d=4$, the organizing matrices can be constructed  from generators in an $n$-dimensional irreducible representation
of $\mathfrak{sl}_2$. The generators chosen for the $d=4$ case \eqref{4dJn} differ from those used in the $d=2$ case \eqref{2dJn}. However, the two sets of generators can be related by simple transformations involving the $n \times n$ diagonal matrix of index values: 
\begin{align}
N=\operatorname{diag}(1,2,\ldots,n) \ , ~~~
J_0\big|_{\eqref{4dJn}}=J_0\big|_{\eqref{2dJn}} \ , ~~
J_-\big|_{\eqref{4dJn}}=-NJ_-\big|_{\eqref{2dJn}}\ , ~~
J_+\big|_{\eqref{4dJn}}=-J_+\big|_{\eqref{2dJn}} N^{-1} \ . \nn\\
\end{align} 
In \eqref{4dMJ}, we have absorbed the factors of $N$ into the generators, resulting in a simpler expression.

\item
The $d=2$ matrix is first order in $\d_t$, with the parameter $\eta$ entering through the term proportional to $J_+$. On the other hand, the $d=4$ matrix contains a product of derivatives involving both $\d_t$ and $\partial_{\bar t}$, while the parameter $\mu$ contributes through a quadratic generator.  Both matrices depend
linearly on their respective heavy-state background parameters, but their dependence on the $\mathfrak{sl}_2$ generators is different. 

\item 
As discussed in section \ref{sec:2d}, the $d=2$ matrix can
be expressed as a scalar shift of a single $\mathfrak{sl}_2$ element, $X=J_0+J_- -6\eta J_+$. The spectrum can be determined using the fundamental two-dimensional  representation. In contrast, the $d=4$ matrix involves quadratic combinations of the generators, which belong to the universal enveloping algebra $U(\mathfrak{sl}_2)$, beyond $\mathfrak{sl}_2$ itself. In particular, $J_+^2$ vanishes in the fundamental representation but
is generally nonzero in higher-dimensional representations. 
The fundamental representation alone cannot capture the full $d=4$
matrix structure, and the simplified spectral argument based on conjugating a single $\mathfrak{sl}_2$ element does not extend to the $d=4$ case.

\end{enumerate}

The algebraic structure of the matrix also raises an intriguing question about the underlying symmetry. In $d=2$, the differential equations correspond to the large-$c$ limit of the BPZ null-state equations, which are directly tied to Virasoro symmetry. In $d=4$, the richer matrix structure, particularly the presence of the $\mu J_+^2$ term, whose coefficient depends on the $d=4$ central charge, 
invites the possibility of a larger algebraic framework. Whether the $d=4$ matrix can be understood in terms of a known extension of Virasoro
symmetry, or points to a new symmetry in higher dimensions,  remains an open question for future work.

\subsection{\selectfont\titlefontmedium Minimal-Twist Matrix} 

If we focus on the differential equations for minimal-twist operators, the first three equations are given in \eqref{4dFeq1}, \eqref{4dlightconeDelta2}, and \eqref{4dlightconeDelta3}, respectively. To construct the organizing matrix, it is convenient to work with the mixed coordinates $(t,\bar z)$, rather than $(t,\bar t)$; see \eqref{zt}. This choice keeps the lightcone coordinate $\bar z$ explicit and also simplifies the subsequent analysis.

An interesting feature of the null-state-type equations near the lightcone is their overall derivative structure, as illustrated by \eqref{4dQeq1}. This feature persists at higher levels, and allows us to reduce the order of the equations by imposing boundary conditions via the near-lightcone expansion; see the discussion around
\eqref{lightconeF}. 
Denote the corresponding  differential equations by
\begin{equation}
\mathscr{D}_n V(t,\bar z)\big|_{\Delta=2-n, \tau_{\rm min}}=0  \ , ~~~~~~~ ~ \mathscr{D}_n\big|_{\tau_{\rm min}}= \det {M}^{(n)}\big|_{\tau_{\rm min}} \ , ~~~~~~~~ n=3, 4, 5, \dots
\end{equation} 
We find the corresponding $n\times n$ matrix can be written as  
\begin{equation}
\label{lightconeM}
(M_n)_{ij}\big|_{\tau_{\rm min}}=
\begin{cases}
\big(\d_t+i-1\big)\partial_{\bar z} &~~~~~~  i=j \quad \quad i=3,6,9,\ldots \\
\d_t+i-1 &~~~~~~  i=j \quad \quad i\ne 3,6,9,\ldots\\
1 &~~~~~~  i = j-1 \\
\displaystyle \frac{\mu}{4}(i-2)(i-1)(n+1-i)(n+2-i) &~~~~~~ i=j+2 \\
0 &~~~~~~  otherwise 
\end{cases}
\end{equation}  
where $i,j=1,\ldots,n$. As an example, the matrix at $n=7$,
corresponding to $\Delta=-5$, takes the form
\begin{align}
(M_7)_{ij}\big|_{\tau_{\rm min}}=\left(
\begin{array}{ccccccc}
 \d_t & 1 & 0 & 0 & 0 & 0 & 0 \\
 0 & \d_t+1 & 1 & 0 & 0 & 0 & 0 \\
 15 \mu  & 0 & \left(\d_t+2\right) \partial_{\zb} & 1 & 0 & 0 & 0 \\
 0 & 30 \mu  & 0 & \d_t+3 & 1 & 0 & 0 \\
 0 & 0 & 36 \mu  & 0 & \d_t+4 & 1 & 0 \\
 0 & 0 & 0 & 30 \mu  & 0 & \left(\d_t+5\right) \partial_{\zb} & 1 \\
 0 & 0 & 0 & 0 & 15 \mu  & 0 & \d_t+6 \\
\end{array}
\right)
\end{align} 
The matrix \eqref{lightconeM} reproduces the equations \eqref{4dFeq1}, \eqref{4dlightconeDelta2}, and
\eqref{4dlightconeDelta3} at $n=3,4,5$, respectively, when written in the coordinates $(t,\bar z)$. Since 
the reduction in differential order is achieved by imposing boundary conditions in the near-lightcone expansion, the
minimal-twist matrix is not obtained simply by taking a small-$\bar z$ limit of the full $d=4$ matrix \eqref{4dmatricDn}.

Since the minimal-twist matrix \eqref{lightconeM} is a reduced description of the full matrix \eqref{4dmatricDn}, we do not repeat the analysis of its recursion relation or the product form of its determinant. Nevertheless, its algebraic structure is worth examining. We find that the minimal-twist matrix can be expressed in the following form:
\begin{align}
M_n \big|_{\tau_{\rm min}}
={}&\Bigg[\left(\partial_t+\frac{n-1}{2}\right)I_n+J_0\Bigg] \Bigg[I_n+(\partial_{\bar z}-1)\Pi_n\Bigg] +J_-+\frac{\mu}{4}J_+^2 
\end{align}
where we use the generators defined in \eqref{2dJn} to have a more economical expression, and we introduce an
$n\times n$ diagonal matrix $\Pi_n$, whose $i$-th diagonal entry is $1$ when $i$ is divisible by $3$, and $0$ otherwise:
\begin{align}
(\Pi_n)_{ij}
=\delta_{ij}
\begin{cases}
1, & i=3,6,9,\ldots \\
0, & otherwise 
\end{cases}
\end{align}
For example,
\begin{align}
\Pi_3&=
\begin{pmatrix}
0&0&0\\
0&0&0\\
0&0&1
\end{pmatrix}
&
\Pi_4&=
\begin{pmatrix}
0&0&0&0\\
0&0&0&0\\
0&0&1&0\\
0&0&0&0
\end{pmatrix}
\\[5pt]
\Pi_5&=
\begin{pmatrix}
0&0&0&0&0\\
0&0&0&0&0\\
0&0&1&0&0\\
0&0&0&0&0\\
0&0&0&0&0
\end{pmatrix}
&
\Pi_6&=
\begin{pmatrix}
0&0&0&0&0&0\\
0&0&0&0&0&0\\
0&0&1&0&0&0\\
0&0&0&0&0&0\\
0&0&0&0&0&0\\
0&0&0&0&0&1
\end{pmatrix}
\end{align}
Each time $n$ increases by three, an additional $1$ appears on the diagonal. 
The multiplication by $\Pi_n$ acts like a projector, which also explains the pattern discontinuity, noticed previously in \cite{Huang:2026jbv}, between the first three $d=4$ near-lightcone equations at $n=3,4,5$ and the differential equation at $n=6$.  

The $d=4$ minimal-twist matrix \eqref{lightconeM} and the $d=2$ organizing matrix \eqref{2dMJ}, $i.e.$, $M=\left(\partial_t+\frac{n-1}{2}\right)I_n+J_0+J_--6\eta J_+ $, 
share a similar, though not identical, algebraic structure. Both can be expressed in terms of generators in the same 
$n$-dimensional representation of $\mathfrak{sl}_2$.  
In the $d=2$ case, the generators enter linearly, and the matrix is a scalar shift of a single $\mathfrak{sl}_2$ element. The $d=4$ minimal-twist matrix
contains an additional dependence on the derivative $\partial_{\bar z}$ through the period-three modulation of its diagonal
entries, and includes the quadratic generator term $\mu J_+^2$ that depends on the central charge.  
 
\section{\selectfont\titlefontmedium Discussion}
\label{sec:dis}

The main result of this work is to show that the null-state-type CFT differential equations, in both $d=2$ and $d=4$, derived from dual gravity can be encapsulated using finite-dimensional matrices built from $\mathfrak{sl}_2$ generators, with heavy-state background deformations encoding the central-charge dependence.

Several directions merit further investigation. 

\begin{itemize}
    \setlength{\itemsep}{3pt}
    \setlength{\parskip}{0pt}

 \item   Beyond their computational efficiency, the matrix structure  in $d=4$ raises a question: can it be understood within a larger algebraic framework? The appearance of the central-charge dependent term involving a quadratic $\mathfrak{sl}_2$-generators in \eqref{4dMJ} is suggestive, but further evidence is needed to establish whether it reflects a higher-dimensional symmetry. It would be interesting to derive the $d=4$ organizing matrix directly from a large-$N$ CFT framework.

  \item  Another direction is to investigate how the $d=4$ matrix is modified by higher-curvature bulk corrections, while keeping the central charge  infinitely large. Two complementary cases are worth distinguishing. In the lightcone limit, for a minimally-coupled bulk scalar, the contributions from minimal-twist multi-stress tensor operators are universal, insensitive to higher-curvature corrections \cite{Fitzpatrick:2019zqz}. Probing the corrections in the stress-tensor sector near the lightcone therefore requires non-minimal bulk interactions; see \cite{Fitzpatrick:2020yjb} for some explicit examples. Moving away from the lightcone, on the other hand, higher-derivative bulk corrections can modify the conformal correlator even assuming a minimally-coupled bulk scalar. Constructing the corresponding matrices in these cases could provide a systematic way to organize the corrections to the CFT differential equations and their algebraic structures. 

 \item  In this work, we have focused on CFTs in $d=2$ and $d=4$. We expect a similar matrix organization can be developed in other dimensions. In particular, although conformal blocks have simpler expressions in even dimensions, it would be useful to investigate the corresponding null-state-type differential equations at large central charge and find the organizing matrices in odd dimensions.

 \item The light scalars associated with these null-state-type differential equations have conformal dimensions outside the unitary regime. Since conformal blocks are meromorphic functions of the conformal dimensions, results obtained at non-unitary values may provide information about correlators at more general conformal dimensions through analytic continuation. For instance, \cite{Chen:2016cms} discusses how the $d=2$ degenerate scalar correlators at non-unitary dimensions can be analytically continued to extract general information about the large-$c$ Virasoro vacuum block. It would be interesting to extend this approach to higher dimensions by exploiting degenerate-type operators and their associated differential equations.

\end{itemize}

It would also be interesting to investigate whether these CFT differential equations encode signatures of the black hole singularity, for example, through the analytic structure of their solutions.  

A related and challenging question is whether the null-state-type equations derived from gravity admit a finite-$N$ generalization in spacetime dimensions greater than two.

\section*{\selectfont\titlefontmedium Acknowledgements}
  
I would like to thank Chris Herzog for related comments.   
This work was supported in part by TSMC Talent Fund Donation (FD114008-016).   
I acknowledge the use of ChatGPT-6 to improve the editing, cross-check some calculations and claims.   
All results were verified by the author, who takes responsibility for the manuscript.  

\vspace{0.3cm}

\begingroup
\renewcommand{\baselinestretch}{0.985}\selectfont
\setlength{\itemsep}{0pt}
\setlength{\parskip}{0pt}

\bibliographystyle{utphys}
\bibliography{nullMatrix}

\providecommand{\href}[2]{#2}\begingroup\raggedright\begin{thebibliography}{100}

\bibitem{Polyakov:2003interview}
``{The Physics of Scale: Interview with Alexander Polyakov}.'' {History of
  Recent Science and Technology Project}, Feb., 2003.
\newblock
  \url{https://wayback.archive-it.org/9060/20230418132451/https://authors.library.caltech.edu/5456/1/hrst.mit.edu//hrs/renormalization/Polyakov/index.html}.
  Interview recorded in Princeton, New Jersey, on 6 February 2003.

\bibitem{Maldacena:1997re}
J.~M. Maldacena, ``{The Large N limit of superconformal field theories and
  supergravity},'' \href{http://dx.doi.org/10.4310/ATMP.1998.v2.n2.a1}{{\em
  Adv. Theor. Math. Phys.} {\bfseries 2} (1998) 231--252},
  \href{http://arxiv.org/abs/hep-th/9711200}{{\ttfamily arXiv:hep-th/9711200}}.

\bibitem{Gubser:1998bc}
S.~S. Gubser, I.~R. Klebanov, and A.~M. Polyakov, ``{Gauge theory correlators
  from noncritical string theory},''
  \href{http://dx.doi.org/10.1016/S0370-2693(98)00377-3}{{\em Phys. Lett. B}
  {\bfseries 428} (1998) 105--114},
  \href{http://arxiv.org/abs/hep-th/9802109}{{\ttfamily arXiv:hep-th/9802109}}.

\bibitem{Witten:1998qj}
E.~Witten, ``{Anti-de Sitter space and holography},''
  \href{http://dx.doi.org/10.4310/ATMP.1998.v2.n2.a2}{{\em Adv. Theor. Math.
  Phys.} {\bfseries 2} (1998) 253--291},
  \href{http://arxiv.org/abs/hep-th/9802150}{{\ttfamily arXiv:hep-th/9802150}}.

\bibitem{Poland:2018epd}
D.~Poland, S.~Rychkov, and A.~Vichi, ``{The Conformal Bootstrap: Theory,
  Numerical Techniques, and Applications},''
  \href{http://dx.doi.org/10.1103/RevModPhys.91.015002}{{\em Rev. Mod. Phys.}
  {\bfseries 91} (2019) 015002},
\href{http://arxiv.org/abs/1805.04405}{{\ttfamily arXiv:1805.04405 [hep-th]}}.

\bibitem{BELAVIN1984333}
A.~Belavin, A.~Polyakov, and A.~Zamolodchikov, ``Infinite conformal symmetry in
  two-dimensional quantum field theory,''
  \href{http://dx.doi.org/https://doi.org/10.1016/0550-3213(84)90052-X}{{\em
  Nuclear Physics B} {\bfseries 241} no.~2, (1984) 333--380}.

\bibitem{Osborn:1993cr}
H.~Osborn and A.~C. Petkou, ``{Implications of conformal invariance in field
  theories for general dimensions},''
  \href{http://dx.doi.org/10.1006/aphy.1994.1045}{{\em Annals Phys.} {\bfseries
  231} (1994) 311--362}, \href{http://arxiv.org/abs/hep-th/9307010}{{\ttfamily
  arXiv:hep-th/9307010}}.

\bibitem{zamolodchikov1984conformal}
A.~B. Zamolodchikov, ``Conformal symmetry in two dimensions: An explicit
  recurrence formula for the conformal partial wave amplitude,''
  \href{http://dx.doi.org/10.1007/BF01214585}{{\em Commun. Math. Phys.}
  {\bfseries 96} no.~3, (1984) 419--422}.

\bibitem{zamolodchikov2}
A.~B. Zamolodchikov, ``Conformal symmetry in two-dimensional space: Recursion
  representation of conformal block,''
  \href{http://dx.doi.org/10.1007/BF01022967}{{\em Theor. Math. Phys.}
  {\bfseries 73} no.~1, (1987) 1088--1093}.

\bibitem{Harlow:2011ny}
D.~Harlow, J.~Maltz, and E.~Witten, ``{Analytic Continuation of Liouville
  Theory},'' \href{http://dx.doi.org/10.1007/JHEP12(2011)071}{{\em JHEP}
  {\bfseries 12} (2011) 071}, \href{http://arxiv.org/abs/1108.4417}{{\ttfamily
  arXiv:1108.4417 [hep-th]}}.

\bibitem{Hartman:2013mia}
T.~Hartman, ``{Entanglement Entropy at Large Central Charge},''
  \href{http://arxiv.org/abs/1303.6955}{{\ttfamily arXiv:1303.6955 [hep-th]}}.

\bibitem{Fitzpatrick:2014vua}
A.~L. Fitzpatrick, J.~Kaplan, and M.~T. Walters, ``{Universality of
  Long-Distance AdS Physics from the CFT Bootstrap},''
  \href{http://dx.doi.org/10.1007/JHEP08(2014)145}{{\em JHEP} {\bfseries 08}
  (2014) 145}, \href{http://arxiv.org/abs/1403.6829}{{\ttfamily arXiv:1403.6829
  [hep-th]}}.

\bibitem{Perlmutter:2015iya}
E.~Perlmutter, ``{Virasoro conformal blocks in closed form},''
  \href{http://dx.doi.org/10.1007/JHEP08(2015)088}{{\em JHEP} {\bfseries 08}
  (2015) 088}, \href{http://arxiv.org/abs/1502.07742}{{\ttfamily
  arXiv:1502.07742 [hep-th]}}.

\bibitem{Hijano:2015qja}
E.~Hijano, P.~Kraus, E.~Perlmutter, and R.~Snively, ``{Semiclassical Virasoro
  blocks from AdS$_{3}$ gravity},''
  \href{http://dx.doi.org/10.1007/JHEP12(2015)077}{{\em JHEP} {\bfseries 12}
  (2015) 077}, \href{http://arxiv.org/abs/1508.04987}{{\ttfamily
  arXiv:1508.04987 [hep-th]}}.

\bibitem{Fitzpatrick:2015foa}
A.~L. Fitzpatrick, J.~Kaplan, M.~T. Walters, and J.~Wang, ``{Hawking from
  Catalan},'' \href{http://dx.doi.org/10.1007/JHEP05(2016)069}{{\em JHEP}
  {\bfseries 05} (2016) 069}, \href{http://arxiv.org/abs/1510.00014}{{\ttfamily
  arXiv:1510.00014 [hep-th]}}.

\bibitem{Beccaria:2015shq}
M.~Beccaria, A.~Fachechi, and G.~Macorini, ``{Virasoro vacuum block at
  next-to-leading order in the heavy-light limit},''
  \href{http://dx.doi.org/10.1007/JHEP02(2016)072}{{\em JHEP} {\bfseries 02}
  (2016) 072}, \href{http://arxiv.org/abs/1511.05452}{{\ttfamily
  arXiv:1511.05452 [hep-th]}}.

\bibitem{Fitzpatrick:2015dlt}
A.~L. Fitzpatrick and J.~Kaplan, ``{Conformal Blocks Beyond the Semi-Classical
  Limit},'' \href{http://dx.doi.org/10.1007/JHEP05(2016)075}{{\em JHEP}
  {\bfseries 05} (2016) 075}, \href{http://arxiv.org/abs/1512.03052}{{\ttfamily
  arXiv:1512.03052 [hep-th]}}.

\bibitem{Fitzpatrick:2016ive}
A.~L. Fitzpatrick, J.~Kaplan, D.~Li, and J.~Wang, ``{On information loss in
  AdS$_{3}$/CFT$_{2}$},'' \href{http://dx.doi.org/10.1007/JHEP05(2016)109}{{\em
  JHEP} {\bfseries 05} (2016) 109},
  \href{http://arxiv.org/abs/1603.08925}{{\ttfamily arXiv:1603.08925
  [hep-th]}}.

\bibitem{Chen:2016cms}
H.~Chen, A.~L. Fitzpatrick, J.~Kaplan, D.~Li, and J.~Wang, ``{Degenerate
  Operators and the $1/c$ Expansion: Lorentzian Resummations, High Order
  Computations, and Super-Virasoro Blocks},''
  \href{http://dx.doi.org/10.1007/JHEP03(2017)167}{{\em JHEP} {\bfseries 03}
  (2017) 167}, \href{http://arxiv.org/abs/1606.02659}{{\ttfamily
  arXiv:1606.02659 [hep-th]}}.

\bibitem{Fitzpatrick:2016mtp}
A.~L. Fitzpatrick, J.~Kaplan, D.~Li, and J.~Wang, ``{Exact Virasoro Blocks from
  Wilson Lines and Background-Independent Operators},''
  \href{http://dx.doi.org/10.1007/JHEP07(2017)092}{{\em JHEP} {\bfseries 07}
  (2017) 092}, \href{http://arxiv.org/abs/1612.06385}{{\ttfamily
  arXiv:1612.06385 [hep-th]}}.

\bibitem{Hikida:2018dxe}
Y.~Hikida and T.~Uetoko, ``{Conformal blocks from Wilson lines with loop
  corrections},'' \href{http://dx.doi.org/10.1103/PhysRevD.97.086014}{{\em
  Phys. Rev. D} {\bfseries 97} no.~8, (2018) 086014},
  \href{http://arxiv.org/abs/1801.08549}{{\ttfamily arXiv:1801.08549
  [hep-th]}}.

\bibitem{Bombini:2018jrg}
A.~Bombini, S.~Giusto, and R.~Russo, ``{A note on the Virasoro blocks at order
  $1/c$},'' \href{http://dx.doi.org/10.1140/epjc/s10052-018-6522-5}{{\em Eur.
  Phys. J. C} {\bfseries 79} no.~1, (2019) 3},
  \href{http://arxiv.org/abs/1807.07886}{{\ttfamily arXiv:1807.07886
  [hep-th]}}.

\bibitem{Cotler:2019nbi}
J.~Cotler, K.~Jensen, and A.~Maloney, ``{Low-dimensional de Sitter quantum
  gravity},'' \href{http://dx.doi.org/10.1007/JHEP06(2020)048}{{\em JHEP}
  {\bfseries 06} (2020) 048}, \href{http://arxiv.org/abs/1905.03780}{{\ttfamily
  arXiv:1905.03780 [hep-th]}}.

\bibitem{Besken:2019jyw}
M.~Be\c{s}ken, S.~Datta, and P.~Kraus, ``{Semi-classical Virasoro blocks: proof
  of exponentiation},'' \href{http://dx.doi.org/10.1007/JHEP01(2020)109}{{\em
  JHEP} {\bfseries 01} (2020) 109},
  \href{http://arxiv.org/abs/1910.04169}{{\ttfamily arXiv:1910.04169
  [hep-th]}}.

\bibitem{Anous:2020vtw}
T.~Anous and F.~M. Haehl, ``{On the Virasoro six-point identity block and
  chaos},'' \href{http://dx.doi.org/10.1007/JHEP08(2020)002}{{\em JHEP}
  {\bfseries 08} no.~08, (2020) 002},
  \href{http://arxiv.org/abs/2005.06440}{{\ttfamily arXiv:2005.06440
  [hep-th]}}.

\bibitem{Nguyen:2021jja}
K.~Nguyen, ``{Reparametrization modes in 2d CFT and the effective theory of
  stress tensor exchanges},''
  \href{http://dx.doi.org/10.1007/JHEP05(2021)029}{{\em JHEP} {\bfseries 05}
  (2021) 029}, \href{http://arxiv.org/abs/2101.08800}{{\ttfamily
  arXiv:2101.08800 [hep-th]}}.

\bibitem{Pavlov:2024zqj}
M.~Pavlov, ``{Differential equations for classical Virasoro blocks with heavy
  and light operators},''
  \href{http://dx.doi.org/10.1140/epjc/s10052-025-14688-8}{{\em Eur. Phys. J.
  C} {\bfseries 85} no.~9, (2025) 982},
  \href{http://arxiv.org/abs/2408.15967}{{\ttfamily arXiv:2408.15967
  [hep-th]}}.

\bibitem{Bissi:2024wur}
A.~Bissi, N.~Dondi, A.~Piazza, T.~Reis, and M.~Serone, ``{On the 1/c expansion
  in 2d CFTs with degenerate operators},''
  \href{http://dx.doi.org/10.1007/JHEP03(2025)092}{{\em JHEP} {\bfseries 03}
  (2025) 092}, \href{http://arxiv.org/abs/2412.04387}{{\ttfamily
  arXiv:2412.04387 [hep-th]}}.

\bibitem{Fortin:2025asz}
J.-F. Fortin, L.~Quintavalle, and W.~Skiba, ``{Virasoro OPE and conformal
  blocks from the inverse Shapovalov form},''
  \href{http://dx.doi.org/10.21468/SciPostPhys.20.5.140}{{\em SciPost Phys.}
  {\bfseries 20} no.~5, (2026) 140},
  \href{http://arxiv.org/abs/2509.09765}{{\ttfamily arXiv:2509.09765
  [hep-th]}}.

\bibitem{Ceplak:2024bja}
N.~{\v{C}}eplak, H.~Liu, A.~Parnachev, and S.~Valach, ``{Black hole singularity
  from OPE},'' \href{http://dx.doi.org/10.1007/JHEP10(2024)105}{{\em JHEP}
  {\bfseries 10} (2024) 105}, \href{http://arxiv.org/abs/2404.17286}{{\ttfamily
  arXiv:2404.17286 [hep-th]}}.

\bibitem{Kraus:2002iv}
P.~Kraus, H.~Ooguri, and S.~Shenker, ``{Inside the horizon with AdS/CFT},''
  \href{http://dx.doi.org/10.1103/PhysRevD.67.124022}{{\em Phys. Rev. D}
  {\bfseries 67} (2003) 124022},
  \href{http://arxiv.org/abs/hep-th/0212277}{{\ttfamily arXiv:hep-th/0212277}}.

\bibitem{Fidkowski:2003nf}
L.~Fidkowski, V.~Hubeny, M.~Kleban, and S.~Shenker, ``{The Black hole
  singularity in AdS/CFT},''
  \href{http://dx.doi.org/10.1088/1126-6708/2004/02/014}{{\em JHEP} {\bfseries
  02} (2004) 014}, \href{http://arxiv.org/abs/hep-th/0306170}{{\ttfamily
  arXiv:hep-th/0306170}}.

\bibitem{Festuccia:2005pi}
G.~Festuccia and H.~Liu, ``{Excursions beyond the horizon: Black hole
  singularities in Yang-Mills theories. I.},''
  \href{http://dx.doi.org/10.1088/1126-6708/2006/04/044}{{\em JHEP} {\bfseries
  04} (2006) 044}, \href{http://arxiv.org/abs/hep-th/0506202}{{\ttfamily
  arXiv:hep-th/0506202}}.

\bibitem{Iliesiu:2018fao}
L.~Iliesiu, M.~Kolo{\u{g}}lu, R.~Mahajan, E.~Perlmutter, and D.~Simmons-Duffin,
  ``{The Conformal Bootstrap at Finite Temperature},''
  \href{http://dx.doi.org/10.1007/JHEP10(2018)070}{{\em JHEP} {\bfseries 10}
  (2018) 070}, \href{http://arxiv.org/abs/1802.10266}{{\ttfamily
  arXiv:1802.10266 [hep-th]}}.

\bibitem{Gobeil:2018fzy}
Y.~Gobeil, A.~Maloney, G.~S. Ng, and J.-q. Wu, ``{Thermal Conformal Blocks},''
  \href{http://dx.doi.org/10.21468/SciPostPhys.7.2.015}{{\em SciPost Phys.}
  {\bfseries 7} no.~2, (2019) 015},
  \href{http://arxiv.org/abs/1802.10537}{{\ttfamily arXiv:1802.10537
  [hep-th]}}.

\bibitem{Kulaxizi:2018dxo}
M.~Kulaxizi, G.~S. Ng, and A.~Parnachev, ``{Black Holes, Heavy States, Phase
  Shift and Anomalous Dimensions},''
  \href{http://dx.doi.org/10.21468/SciPostPhys.6.6.065}{{\em SciPost Phys.}
  {\bfseries 6} no.~6, (2019) 065},
  \href{http://arxiv.org/abs/1812.03120}{{\ttfamily arXiv:1812.03120
  [hep-th]}}.

\bibitem{Fitzpatrick:2019zqz}
A.~L. Fitzpatrick and K.-W. Huang, ``{Universal Lowest-Twist in CFTs from
  Holography},'' \href{http://dx.doi.org/10.1007/JHEP08(2019)138}{{\em JHEP}
  {\bfseries 08} (2019) 138}, \href{http://arxiv.org/abs/1903.05306}{{\ttfamily
  arXiv:1903.05306 [hep-th]}}.

\bibitem{Li:2019tpf}
Y.-Z. Li, Z.-F. Mai, and H.~L{\"u}, ``{Holographic OPE Coefficients from AdS
  Black Holes with Matters},''
  \href{http://dx.doi.org/10.1007/JHEP09(2019)001}{{\em JHEP} {\bfseries 09}
  (2019) 001}, \href{http://arxiv.org/abs/1905.09302}{{\ttfamily
  arXiv:1905.09302 [hep-th]}}.

\bibitem{Kulaxizi:2019tkd}
M.~Kulaxizi, G.~S. Ng, and A.~Parnachev, ``{Subleading Eikonal, AdS/CFT and
  Double Stress Tensors},''
  \href{http://dx.doi.org/10.1007/JHEP10(2019)107}{{\em JHEP} {\bfseries 10}
  (2019) 107},
\href{http://arxiv.org/abs/1907.00867}{{\ttfamily arXiv:1907.00867 [hep-th]}}.

\bibitem{Fitzpatrick:2019efk}
A.~L. Fitzpatrick, K.-W. Huang, and D.~Li, ``{Probing universalities in d
  {\ensuremath{>}} 2 CFTs: from black holes to shockwaves},''
  \href{http://dx.doi.org/10.1007/JHEP11(2019)139}{{\em JHEP} {\bfseries 11}
  (2019) 139}, \href{http://arxiv.org/abs/1907.10810}{{\ttfamily
  arXiv:1907.10810 [hep-th]}}.

\bibitem{Karlsson:2019dbd}
R.~Karlsson, M.~Kulaxizi, A.~Parnachev, and P.~Tadić, ``{Leading Multi-Stress
  Tensors and Conformal Bootstrap},''
  \href{http://dx.doi.org/10.1007/JHEP01(2020)076}{{\em JHEP} {\bfseries 01}
  (2020) 076},
\href{http://arxiv.org/abs/1909.05775}{{\ttfamily arXiv:1909.05775 [hep-th]}}.

\bibitem{Li:2019zba}
Y.-Z. Li, ``{Heavy-light Bootstrap from Lorentzian Inversion Formula},''
  \href{http://dx.doi.org/10.1007/JHEP07(2020)046}{{\em JHEP} {\bfseries 07}
  (2020) 046}, \href{http://arxiv.org/abs/1910.06357}{{\ttfamily
  arXiv:1910.06357 [hep-th]}}.

\bibitem{Karlsson:2019txu}
R.~Karlsson, ``{Multi-stress tensors and next-to-leading singularities in the
  Regge limit},'' \href{http://dx.doi.org/10.1007/JHEP08(2020)037}{{\em JHEP}
  {\bfseries 08} (2020) 037}, \href{http://arxiv.org/abs/1912.01577}{{\ttfamily
  arXiv:1912.01577 [hep-th]}}.

\bibitem{Karlsson:2020ghx}
R.~Karlsson, M.~Kulaxizi, A.~Parnachev, and P.~Tadi\'c, ``{Stress tensor sector
  of conformal correlators operators in the Regge limit},''
  \href{http://dx.doi.org/10.1007/JHEP07(2020)019}{{\em JHEP} {\bfseries 07}
  (2020) 019}, \href{http://arxiv.org/abs/2002.12254}{{\ttfamily
  arXiv:2002.12254 [hep-th]}}.

\bibitem{Fitzpatrick:2020yjb}
A.~L. Fitzpatrick, K.-W. Huang, D.~Meltzer, E.~Perlmutter, and
  D.~Simmons-Duffin, ``{Model-dependence of minimal-twist OPEs in d
  {\ensuremath{>}} 2 holographic CFTs},''
  \href{http://dx.doi.org/10.1007/JHEP11(2020)060}{{\em JHEP} {\bfseries 11}
  (2020) 060}, \href{http://arxiv.org/abs/2007.07382}{{\ttfamily
  arXiv:2007.07382 [hep-th]}}.

\bibitem{Alday:2020eua}
L.~F. Alday, M.~Kologlu, and A.~Zhiboedov, ``{Holographic correlators at finite
  temperature},'' \href{http://dx.doi.org/10.1007/JHEP06(2021)082}{{\em JHEP}
  {\bfseries 06} (2021) 082}, \href{http://arxiv.org/abs/2009.10062}{{\ttfamily
  arXiv:2009.10062 [hep-th]}}.

\bibitem{Grinberg:2020fdj}
M.~Grinberg and J.~Maldacena, ``{Proper time to the black hole singularity from
  thermal one-point functions},''
  \href{http://dx.doi.org/10.1007/JHEP03(2021)131}{{\em JHEP} {\bfseries 03}
  (2021) 131}, \href{http://arxiv.org/abs/2011.01004}{{\ttfamily
  arXiv:2011.01004 [hep-th]}}.

\bibitem{Karlsson:2021duj}
R.~Karlsson, A.~Parnachev, and P.~Tadi{\'c}, ``{Thermalization in large-N
  CFTs},'' \href{http://dx.doi.org/10.1007/JHEP09(2021)205}{{\em JHEP}
  {\bfseries 09} (2021) 205}, \href{http://arxiv.org/abs/2102.04953}{{\ttfamily
  arXiv:2102.04953 [hep-th]}}.

\bibitem{Rodriguez-Gomez:2021mkk}
D.~Rodriguez-Gomez and J.~G. Russo, ``{Thermal correlation functions in CFT and
  factorization},'' \href{http://dx.doi.org/10.1007/JHEP11(2021)049}{{\em JHEP}
  {\bfseries 11} (2021) 049}, \href{http://arxiv.org/abs/2105.13909}{{\ttfamily
  arXiv:2105.13909 [hep-th]}}.

\bibitem{Krishna:2021fus}
H.~Krishna and D.~Rodriguez-Gomez, ``{Holographic thermal correlators
  revisited},'' \href{http://dx.doi.org/10.1007/JHEP11(2021)139}{{\em JHEP}
  {\bfseries 11} (2021) 139}, \href{http://arxiv.org/abs/2108.00277}{{\ttfamily
  arXiv:2108.00277 [hep-th]}}.

\bibitem{Karlsson:2022osn}
R.~Karlsson, A.~Parnachev, V.~Prilepina, and S.~Valach, ``{Thermal stress
  tensor correlators, OPE and holography},''
  \href{http://dx.doi.org/10.1007/JHEP09(2022)234}{{\em JHEP} {\bfseries 09}
  (2022) 234}, \href{http://arxiv.org/abs/2206.05544}{{\ttfamily
  arXiv:2206.05544 [hep-th]}}.

\bibitem{Dodelson:2022yvn}
M.~Dodelson, A.~Grassi, C.~Iossa, D.~Panea~Lichtig, and A.~Zhiboedov,
  ``{Holographic thermal correlators from supersymmetric instantons},''
  \href{http://dx.doi.org/10.21468/SciPostPhys.14.5.116}{{\em SciPost Phys.}
  {\bfseries 14} no.~5, (2023) 116},
  \href{http://arxiv.org/abs/2206.07720}{{\ttfamily arXiv:2206.07720
  [hep-th]}}.

\bibitem{Parisini:2022wkb}
E.~Parisini, K.~Skenderis, and B.~Withers, ``{Embedding formalism for CFTs in
  general states on curved backgrounds},''
  \href{http://dx.doi.org/10.1103/PhysRevD.107.066022}{{\em Phys. Rev. D}
  {\bfseries 107} no.~6, (2023) 066022},
  \href{http://arxiv.org/abs/2209.09250}{{\ttfamily arXiv:2209.09250
  [hep-th]}}.

\bibitem{Dodelson:2023vrw}
M.~Dodelson, C.~Iossa, R.~Karlsson, and A.~Zhiboedov, ``{A thermal product
  formula},'' \href{http://dx.doi.org/10.1007/JHEP01(2024)036}{{\em JHEP}
  {\bfseries 01} (2024) 036}, \href{http://arxiv.org/abs/2304.12339}{{\ttfamily
  arXiv:2304.12339 [hep-th]}}.

\bibitem{Esper:2023jeq}
C.~Esper, K.-W. Huang, R.~Karlsson, A.~Parnachev, and S.~Valach, ``{Thermal
  stress tensor correlators near lightcone and holography},''
  \href{http://dx.doi.org/10.1007/JHEP11(2023)107}{{\em JHEP} {\bfseries 11}
  (2023) 107}, \href{http://arxiv.org/abs/2306.00787}{{\ttfamily
  arXiv:2306.00787 [hep-th]}}.

\bibitem{Parisini:2023nbd}
E.~Parisini, K.~Skenderis, and B.~Withers, ``{The ambient space formalism},''
  \href{http://dx.doi.org/10.1007/JHEP05(2024)296}{{\em JHEP} {\bfseries 05}
  (2024) 296}, \href{http://arxiv.org/abs/2312.03820}{{\ttfamily
  arXiv:2312.03820 [hep-th]}}.

\bibitem{Buric:2025anb}
I.~Buri{\'c}, I.~Gusev, and A.~Parnachev, ``{Thermal holographic correlators
  and KMS condition},'' \href{http://dx.doi.org/10.1007/JHEP09(2025)053}{{\em
  JHEP} {\bfseries 09} (2025) 053},
  \href{http://arxiv.org/abs/2505.10277}{{\ttfamily arXiv:2505.10277
  [hep-th]}}.

\bibitem{Haehl:2025ehf}
F.~M. Haehl and K.-W. Huang, ``{Virasoro OPE blocks, causal diamonds, and
  higher-dimensional CFT},''
  \href{http://dx.doi.org/10.1007/JHEP02(2026)030}{{\em JHEP} {\bfseries 02}
  (2026) 030}, \href{http://arxiv.org/abs/2505.12456}{{\ttfamily
  arXiv:2505.12456 [hep-th]}}.

\bibitem{Buric:2025uqt}
I.~Buri{\'c}, F.~Mangialardi, F.~Russo, V.~Schomerus, and A.~Vichi,
  ``{Heavy-heavy-light asymptotics from thermal correlators},''
  \href{http://dx.doi.org/10.1007/JHEP04(2026)027}{{\em JHEP} {\bfseries 04}
  (2026) 027}, \href{http://arxiv.org/abs/2506.21671}{{\ttfamily
  arXiv:2506.21671 [hep-th]}}.

\bibitem{Ammon:2025cdz}
M.~Ammon, J.~Hollweck, T.~H{\"o}ssel, and K.~W{\"o}lfl, ``{Thermal n-point
  conformal blocks in four dimensions from oscillator representations},''
  \href{http://dx.doi.org/10.1007/JHEP04(2026)102}{{\em JHEP} {\bfseries 04}
  (2026) 102}, \href{http://arxiv.org/abs/2507.22974}{{\ttfamily
  arXiv:2507.22974 [hep-th]}}.

\bibitem{Buric:2025fye}
I.~Buri{\'c}, I.~Gusev, and A.~Parnachev, ``{Holographic correlators from
  thermal bootstrap},'' \href{http://dx.doi.org/10.1007/JHEP05(2026)059}{{\em
  JHEP} {\bfseries 05} (2026) 059},
  \href{http://arxiv.org/abs/2508.08373}{{\ttfamily arXiv:2508.08373
  [hep-th]}}.

\bibitem{Niarchos:2025cdg}
V.~Niarchos, C.~Papageorgakis, A.~Stratoudakis, and M.~Woolley, ``{Deep finite
  temperature bootstrap},'' \href{http://dx.doi.org/10.1103/qrjx-w4md}{{\em
  Phys. Rev. D} {\bfseries 112} no.~12, (2025) 126012},
  \href{http://arxiv.org/abs/2508.08560}{{\ttfamily arXiv:2508.08560
  [hep-th]}}.

\bibitem{Bajc:2025jjv}
B.~Bajc and K.~Trailovi{\'c}, ``{Holographic thermal propagator from
  modularity},'' \href{http://dx.doi.org/10.1007/JHEP11(2025)133}{{\em JHEP}
  {\bfseries 11} (2025) 133}, \href{http://arxiv.org/abs/2509.02226}{{\ttfamily
  arXiv:2509.02226 [hep-th]}}.

\bibitem{Barrat:2025twb}
J.~Barrat, D.~N. Bozkurt, E.~Marchetto, A.~Miscioscia, and E.~Pomoni,
  ``{Analytic thermal bootstrap meets holography},''
  \href{http://dx.doi.org/10.1007/JHEP05(2026)180}{{\em JHEP} {\bfseries 05}
  (2026) 180}, \href{http://arxiv.org/abs/2510.20894}{{\ttfamily
  arXiv:2510.20894 [hep-th]}}.

\bibitem{Afkhami-Jeddi:2025wra}
N.~Afkhami-Jeddi, S.~Caron-Huot, J.~Chakravarty, and A.~Maloney, ``{Imprint of
  the black hole singularity on thermal two-point functions},''
  \href{http://dx.doi.org/10.1103/1j3s-fzjl}{{\em Phys. Rev. D} {\bfseries 114}
  no.~2, (2026) 025005}, \href{http://arxiv.org/abs/2510.21673}{{\ttfamily
  arXiv:2510.21673 [hep-th]}}.

\bibitem{Dodelson:2025jff}
M.~Dodelson, C.~Iossa, and R.~Karlsson, ``{Bouncing off a stringy
  singularity},'' \href{http://dx.doi.org/10.1007/JHEP07(2026)161}{{\em JHEP}
  {\bfseries 07} (2026) 161}, \href{http://arxiv.org/abs/2511.09616}{{\ttfamily
  arXiv:2511.09616 [hep-th]}}.

\bibitem{Ceplak:2025dds}
N.~{\v{C}}eplak, H.~Liu, A.~Parnachev, and S.~Valach, ``{Fooling the censor:
  going beyond inner horizons with the OPE},''
  \href{http://dx.doi.org/10.1007/JHEP05(2026)001}{{\em JHEP} {\bfseries 05}
  (2026) 001}, \href{http://arxiv.org/abs/2511.09638}{{\ttfamily
  arXiv:2511.09638 [hep-th]}}.

\bibitem{Jia:2026pmv}
Y.~Jia and M.~Kulaxizi, ``{Bulk Phase Shift and Singularity},''
  \href{http://arxiv.org/abs/2602.06558}{{\ttfamily arXiv:2602.06558
  [hep-th]}}.

\bibitem{Araya:2026shz}
I.~J. Araya, C.~Esper, Y.~Jia, M.~Kulaxizi, and A.~Parnachev, ``{Bulkcone
  Singularities and Complex Geodesics},''
  \href{http://arxiv.org/abs/2602.12893}{{\ttfamily arXiv:2602.12893
  [hep-th]}}.

\bibitem{Giombi:2026kdz}
S.~Giombi, Y.-Z. Li, and J.~Shan, ``{Bouncing singularities and thermal
  correlators on line defects},''
  \href{http://dx.doi.org/10.1007/JHEP07(2026)170}{{\em JHEP} {\bfseries 07}
  (2026) 170}, \href{http://arxiv.org/abs/2603.11012}{{\ttfamily
  arXiv:2603.11012 [hep-th]}}.

\bibitem{Jia:2026ryl}
H.~F. Jia and M.~Rangamani, ``{Exact holographic thermal spectral functions:
  OPE, non-perturbative corrections, and black hole singularity},''
  \href{http://arxiv.org/abs/2604.10803}{{\ttfamily arXiv:2604.10803
  [hep-th]}}.

\bibitem{Arnaudo:2026der}
P.~Arnaudo and B.~Withers, ``{Analytic structure of holographic thermal
  correlators from Fourier series},''
  \href{http://dx.doi.org/10.1007/JHEP06(2026)205}{{\em JHEP} {\bfseries 06}
  (2026) 205}, \href{http://arxiv.org/abs/2603.13469}{{\ttfamily
  arXiv:2603.13469 [hep-th]}}.

\bibitem{Grozdanov:2026cut}
S.~Grozdanov, S.~Valach, and M.~Vrbica, ``{Bouncing geodesics, black hole
  singularities, and singularities of thermal correlators},''
  \href{http://dx.doi.org/10.1007/JHEP08(2026)022}{{\em JHEP} {\bfseries 08}
  (2026) 022}, \href{http://arxiv.org/abs/2603.15598}{{\ttfamily
  arXiv:2603.15598 [hep-th]}}.

\bibitem{Arnaudo:2026tcy}
P.~Arnaudo and B.~Withers, ``{Bouncing singularities in Schwarzschild: a
  geometric origin of the QNM convergence region},''
  \href{http://arxiv.org/abs/2605.16489}{{\ttfamily arXiv:2605.16489 [gr-qc]}}.

\bibitem{Grozdanov:2026ktq}
S.~Grozdanov, V.~Movrin, and S.~Valach, ``{Bouncing Geodesics, Singularities,
  and the Cavity Thermal Product Formula in Asymptotically Flat and de Sitter
  Black Holes},'' \href{http://arxiv.org/abs/2606.11297}{{\ttfamily
  arXiv:2606.11297 [hep-th]}}.

\bibitem{Niarchos:2026wsw}
V.~Niarchos, C.~Papageorgakis, and A.~Stratoudakis, ``{Thermal Double-Twist
  Data in Holography},'' \href{http://arxiv.org/abs/2606.30806}{{\ttfamily
  arXiv:2606.30806 [hep-th]}}.

\bibitem{Arnaudo:2026axe}
P.~Arnaudo, C.~Iossa, R.~Karlsson, and B.~Withers, ``{OPE = QNM},''
  \href{http://arxiv.org/abs/2607.24909}{{\ttfamily arXiv:2607.24909
  [hep-th]}}.

\bibitem{Hartnoll:2026vhu}
S.~A. Hartnoll and A.~Zhiboedov, ``{Can one hear the shape of a black hole
  singularity?},'' \href{http://arxiv.org/abs/2609.07514}{{\ttfamily
  arXiv:2609.07514 [hep-th]}}.

\bibitem{Ceplak:2026ard}
N.~{\v{C}}eplak and S.~Valach, ``{Where's Kasner? OPE Signature of Kasner
  Exponents},'' \href{http://arxiv.org/abs/2609.23776}{{\ttfamily
  arXiv:2609.23776 [hep-th]}}.

\bibitem{Guo:2026xyl}
Y.~Guo, Z.~Li, and T.~Shen, ``{Thermal Polyakov bootstrap},''
  \href{http://arxiv.org/abs/2609.28627}{{\ttfamily arXiv:2609.28627
  [hep-th]}}.

\bibitem{Komargodski:2026oob}
Z.~Komargodski, ``{Fate of Bouncing Singularities at Finite $G_N$},''
  \href{http://arxiv.org/abs/2609.31309}{{\ttfamily arXiv:2609.31309
  [hep-th]}}.

\bibitem{Batra:2026gbq}
G.~Batra, A.~Levine, and S.~H. Shenker, ``{A backreacted bouncing geodesic in
  2D},'' \href{http://arxiv.org/abs/2609.32071}{{\ttfamily arXiv:2609.32071
  [hep-th]}}.

\bibitem{Correia:2026fnr}
M.~Correia, V.~Gon{\c{c}}alves, and F.~Serrano, ``{Effective Field Theory for
  Holographic Compact Objects},''
  \href{http://arxiv.org/abs/2609.37806}{{\ttfamily arXiv:2609.37806
  [hep-th]}}.

\bibitem{Buric:2026sit}
I.~Buri{\'c}, I.~Gusev, and A.~Parnachev, ``{Stringy resolution of bouncing
  singularities},'' \href{http://arxiv.org/abs/2610.12311}{{\ttfamily
  arXiv:2610.12311 [hep-th]}}.

\bibitem{Bajc:2026wzi}
B.~Bajc, M.~Latifi, and K.~Trailovi{\'c}, ``{Holographic thermal propagator at
  finite energy from modularity and triality},''
  \href{http://arxiv.org/abs/2610.02340}{{\ttfamily arXiv:2610.02340
  [hep-th]}}.

\bibitem{Chakravarty:2026fdm}
J.~Chakravarty and Y.~Zigdon, ``{A signature of the timelike singularity in
  holography},'' \href{http://arxiv.org/abs/2610.30134}{{\ttfamily
  arXiv:2610.30134 [hep-th]}}.

\bibitem{Chen:2026efa}
H.-Y. Chen, Y.~Hikida, and Y.~Koga, ``{Quasinormal modes, complex null
  geodesics and singularity resolution in regular AdS black holes},''
  \href{http://arxiv.org/abs/2610.30915}{{\ttfamily arXiv:2610.30915
  [hep-th]}}.

\bibitem{Almheiri:2026cpv}
A.~Almheiri, M.~Dodelson, and Z.~Komargodski, ``{Quantum gravity on the bulk
  cone},'' \href{http://arxiv.org/abs/2610.31610}{{\ttfamily arXiv:2610.31610
  [hep-th]}}.

\bibitem{Oh:2026dky}
H.~Oh, ``{Kasner Exponents from the Bouncing Singularity},''
  \href{http://arxiv.org/abs/2610.31699}{{\ttfamily arXiv:2610.31699
  [hep-th]}}.

\bibitem{Aniceto:2026vtv}
I.~Aniceto, P.~Kovtun, A.~Ratcliffe, and M.~Spali{\'n}ski, ``{Shear viscosity
  from complex light-cone expansion},''
  \href{http://arxiv.org/abs/2610.32892}{{\ttfamily arXiv:2610.32892
  [hep-th]}}.

\bibitem{Grassi:2026ekz}
A.~Grassi and H.~F. Jia, ``{Resurgence of Black Holes in Thermal OPE},''
  \href{http://arxiv.org/abs/2610.37554}{{\ttfamily arXiv:2610.37554
  [hep-th]}}.

\bibitem{Huang:2026jbv}
K.-W. Huang, ``{Holographic derivation of BPZ-type null-state equations in
  higher dimensional CFTs},'' \href{http://dx.doi.org/10.1103/5pbb-fbb6}{{\em
  Phys. Rev. D} {\bfseries 114} no.~8, (2026) L081902},
  \href{http://arxiv.org/abs/2605.04032}{{\ttfamily arXiv:2605.04032
  [hep-th]}}.

\bibitem{Huang:2023ikg}
K.-W. Huang, ``{Toward null-state equations in d {\ensuremath{>}} 2},''
  \href{http://dx.doi.org/10.1007/JHEP11(2023)203}{{\em JHEP} {\bfseries 11}
  (2023) 203}, \href{http://arxiv.org/abs/2308.03229}{{\ttfamily
  arXiv:2308.03229 [hep-th]}}.

\bibitem{Huang:2024wbq}
K.-W. Huang, ``{Resummation of multistress tensors in higher dimensions},''
  \href{http://dx.doi.org/10.1103/PhysRevD.111.046016}{{\em Phys. Rev. D}
  {\bfseries 111} no.~4, (2025) 046016},
  \href{http://arxiv.org/abs/2406.07458}{{\ttfamily arXiv:2406.07458
  [hep-th]}}.

\bibitem{Bauer:1991ai}
M.~Bauer, P.~Di~Francesco, C.~Itzykson, and J.~B. Zuber, ``{Covariant
  differential equations and singular vectors in Virasoro representations},''
  \href{http://dx.doi.org/10.1016/0550-3213(91)90541-5}{{\em Nucl. Phys. B}
  {\bfseries 362} (1991) 515--562}.

\bibitem{DiFrancesco:1997nk}
P.~Di~Francesco, P.~Mathieu, and D.~Senechal, {\em Conformal Field Theory}.
\newblock Graduate Texts in Contemporary Physics. Springer-Verlag, New York,
  1997.

\bibitem{Keski-Vakkuri:1998gmz}
E.~Keski-Vakkuri, ``{Bulk and boundary dynamics in BTZ black holes},''
  \href{http://dx.doi.org/10.1103/PhysRevD.59.104001}{{\em Phys. Rev. D}
  {\bfseries 59} (1999) 104001},
  \href{http://arxiv.org/abs/hep-th/9808037}{{\ttfamily arXiv:hep-th/9808037}}.

\bibitem{breitenlohner1982stability}
P.~Breitenlohner and D.~Z. Freedman, ``{Stability in Gauged Extended
  Supergravity},'' \href{http://dx.doi.org/10.1016/0003-4916(82)90116-6}{{\em
  Annals Phys.} {\bfseries 144} (1982) 249}.

\bibitem{breitenlohner1982positive}
P.~Breitenlohner and D.~Z. Freedman, ``{Positive Energy in anti-De Sitter
  Backgrounds and Gauged Extended Supergravity},''
  \href{http://dx.doi.org/10.1016/0370-2693(82)90643-8}{{\em Phys. Lett. B}
  {\bfseries 115} (1982) 197--201}.

\bibitem{Dolan:2000ut}
F.~A. Dolan and H.~Osborn, ``{Conformal four point functions and the operator
  product expansion},''
  \href{http://dx.doi.org/10.1016/S0550-3213(01)00013-X}{{\em Nucl. Phys. B}
  {\bfseries 599} (2001) 459--496},
  \href{http://arxiv.org/abs/hep-th/0011040}{{\ttfamily arXiv:hep-th/0011040}}.

\bibitem{Fitzpatrick:2012yx}
A.~L. Fitzpatrick, J.~Kaplan, D.~Poland, and D.~Simmons-Duffin, ``{The Analytic
  Bootstrap and AdS Superhorizon Locality},''
  \href{http://dx.doi.org/10.1007/JHEP12(2013)004}{{\em JHEP} {\bfseries 12}
  (2013) 004}, \href{http://arxiv.org/abs/1212.3616}{{\ttfamily arXiv:1212.3616
  [hep-th]}}.

\bibitem{Komargodski:2012ek}
Z.~Komargodski and A.~Zhiboedov, ``{Convexity and Liberation at Large Spin},''
  \href{http://dx.doi.org/10.1007/JHEP11(2013)140}{{\em JHEP} {\bfseries 11}
  (2013) 140}, \href{http://arxiv.org/abs/1212.4103}{{\ttfamily arXiv:1212.4103
  [hep-th]}}.

\end{thebibliography}\endgroup
\endgroup

\end{document}